\documentclass[numberedappendix]{emulateapj-rtx4}

\usepackage{amsmath}
\usepackage{amssymb}
\usepackage{graphicx}
\usepackage[usenames,dvipsnames]{color}
\usepackage{natbib}
\usepackage[backref,breaklinks,colorlinks,citecolor=blue]{hyperref}
\usepackage[all]{hypcap}

\usepackage{sidecap}

\newcommand{\WMAP}{\textsl{WMAP}}

\newcommand{\wmap}{{\WMAP}}
\newcommand{\Planck}{{\textsl{Planck}}}
\newcommand{\planck}{{\textsl{Planck}}}
\newcommand{\lcdm}{\ensuremath{\Lambda}CDM}

\renewcommand{\ell}{\ensuremath{l}}
\newcommand{\be}{\begin{equation}}
\newcommand{\ee}{\end{equation}}
\newcommand{\beq}{\begin{equation}}
\newcommand{\eeq}{\end{equation}}

\def\ba{\begin{eqnarray}}
\def\ea{\end{eqnarray}}

\newcommand{\barr}{\begin{array}}
\newcommand{\earr}{\end{array}}

\begin{document}

\title{Comparing \textsl{PLANCK} and \wmap:  Maps, Spectra, and Parameters}
\author{
{D. Larson}  \altaffilmark{1},
{J. L. Weiland} \altaffilmark{1},
{G. Hinshaw} \altaffilmark{2},
and {C. L. Bennett} \altaffilmark{1}}

\altaffiltext{1}{{Dept. of Physics \& Astronomy, The Johns Hopkins University, 3400 N. Charles St., Baltimore, MD  21218-2686}}
\altaffiltext{2}{{Dept. of Physics and Astronomy, University of British Columbia, Vancouver, BC  Canada V6T 1Z1}}

\email{{dlarso11@jhu.edu}}

\begin{abstract}
We examine the consistency of the 9 yr \wmap\ data and the first-release \planck\ data.  We specifically compare sky maps, power spectra, and the inferred $\Lambda$ cold dark matter (\lcdm) cosmological parameters. Residual dipoles are seen in the \wmap\ and \planck\ sky map differences, but their amplitudes are consistent within the quoted uncertainties, and they are not large enough to explain the widely noted differences in angular power spectra at higher $l$.  We remove the residual dipoles and use templates to remove residual galactic foregrounds; after doing so, the residual difference maps exhibit a quadrupole and other large-scale systematic structure. We identify this structure as possibly originating from \planck's beam sidelobe pick-up, but note that it appears to have insignificant cosmological impact.  We develop an extension of the internal linear combination technique to find the minimum-variance difference between the \wmap\ and \planck\ sky maps; again we find features that plausibly originate in the \planck\ data.  Lacking access to the \planck\ time-ordered data we cannot further assess these features.  We examine \lcdm\ model fits to the angular power spectra and conclude that the $\sim$2.5\% difference in the spectra at multipoles greater than $l \sim 100$ are significant at the 3--5$\sigma$ level, depending on how beam uncertainties are handled in the data.  We revisit the analysis of \wmap's beam data to address the power spectrum differences and conclude that previously derived uncertainties are robust and cannot explain the power spectrum differences.  In fact, any remaining \wmap\ errors are most likely to exacerbate the difference.  Finally, we examine the consistency of the \lcdm\ parameters inferred from each data set taking into account the fact that both experiments observe the same sky, but cover different multipole ranges, apply different sky masks, and have different noise.  We find that, while individual parameter values agree within the uncertainties, the six parameters taken together are discrepant at the $\sim$6$\sigma$ level, with $\chi^2 = 56$ for 6 degrees of freedom (probability to exceed, PTE = $3 \times 10^{-10})$.   The nature of this discrepancy is explored: of the six parameters, $\chi^2$ is best improved by marginalizing over $\Omega_c h^2$, giving $\chi^2=5.2$ for 5 degrees of freedom.  As an exercise, we find that perturbing the \wmap\ window function by its dominant beam error profile has little effect on $\Omega_c h^2$, while perturbing the \planck\ window function by its corresponding error profile has a much greater effect on $\Omega_c h^2$.
\end{abstract}

\section{Introduction}

The March 2013 release of \planck\ data\footnote{\url{http://pla.esac.esa.int/pla/aio/planckProducts.html}} was another milestone in the fast-paced series of advances in observational cosmology.  In December 2012 the \wmap\ team released an analysis of their full nine-year data set \citep{bennett/etal:2013,hinshaw/etal:2013}.  The existence of two (almost completely) independent full-sky surveys of microwave emission using multiple overlapping frequency bands presents an enormous opportunity for cross-checking results. 

There have been surprisingly few attempts to carefully evaluate the statistical consistency of the \wmap\ and \planck\ data.  A thorough treatment requires accounting for both the similarities and differences between the two experiments: while both observe the same sky, they employ different frequency bands (with some partial overlap), they are sensitive to different multipole ranges, they apply different sky masks, and they have different instrument noise.  While we take all of these effects into account in this paper, we are not the first to assess consistency.  \cite{planck/15:2013}, \cite{planck/16:2013}, and \cite{planck/31:2013} report that the \wmap\ angular power spectrum is roughly 2.5\% brighter than \planck's for multipoles greater than $l \sim 100$ (1.25\% in amplitude).  \citet{kovacs/etal:2013} report that the \wmap\ angular power spectrum is about 2.6\% higher than that of \planck\ ``at a very high significance'' on the same scales, while ``at higher multipole moments there appears to be no significant bias.''   \citet{hazra/shafieloo:2014b} tested consistency of the best-fit \lcdm\ parameters: when they marginalize over the spectrum amplitude they find consistency, but when the amplitudes of each are fixed, they find tension at the $3\sigma$ level. They went on \citep{hazra/shafieloo:2014a} to show that a concordance cosmological model fit to all data is consistent with the \planck\ data at only 2 to 3$\sigma$ confidence level; the \planck\ spectrum had a dearth of power at both low-$l$ and high-$l$ relative to the concordance model.  \citet{spergel/flauger/hlozek:2013} found that the power spectrum from the \planck\ $217 \times 217$ GHz detector set is responsible for some of the tension when compared with other cosmological measurements.

\begin{figure}
\begin{center}
\includegraphics[width=0.49\textwidth]{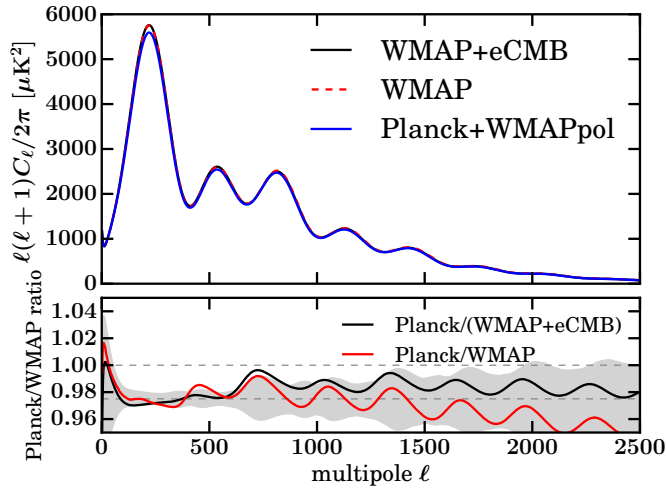}
\end{center}
\caption{A comparison of the best-fit 6-parameter \lcdm\ model spectra derived from recent CMB data sets.  {\em Solid black} - fit to \wmap+eCMB data (SPT and ACT); {\em dashed red} - fit to \wmap\ (only) data; {\em solid blue} - fit to \planck\ plus \wmap\ low-$l$ polarization data.  The bottom panel shows the ratio of the \planck-based fits to the two \wmap-based fits.  The dominant feature is a $\sim$2\% amplitude difference between the fits, though some structure is present in the ratio.   A dashed horizontal line is shown at 0.975 to guide the eye.  The \wmap-only forecast, in red, exhibits an additional $\sim$2\% difference at high $l$, well beyond the $l$ range directly measured by \wmap\ ($l<1200$).  The gray band shows the 1$\sigma$ uncertainty in the ratio accounting for the fact that \wmap\ and \planck\ observe the same sky, but apply different sky masks, cover different multipole ranges, and have independent noise.  Given this uncertainty, we cannot distinguish a simple amplitude rescaling from a more complicated shift in parameters.
\label{fig:planck_wmap_powspec_ratio}}
\end{figure}

Figure~\ref{fig:planck_wmap_powspec_ratio} compares the best-fit \lcdm\ models derived from \planck\ and \wmap\ data.   The top panel shows the models and the bottom panel shows the model ratios. The error band plotted in Figure~\ref{fig:planck_wmap_powspec_ratio} (derived from simulations described in \S\ref{sec:param_compare})  indicates the spread in model fits that would be expected given differences in noise, multipole coverage, and sky masking between \planck\ and \wmap.  The \wmap\ signal is clearly brighter than the \planck\ signal for multipoles  $l \gtrsim 100$.  Given the gray error band, for $l \lesssim 100$ the interpretation of the model fits is less clear.   However, spectrum ratios derived purely from data suggest that the $\sim$2.5\% ratio persists to low $l$, see, for example, Figure~A.1 of \cite{planck/16:2013}.   The range of best-fit models appears to encompass an overall rescaling of the data (and nothing more), but a richer set of parameter shifts is also allowed.  A pure scaling difference would be surprising given that the initial \planck\ data release adopted the \wmap\ dipole signal as an absolute brightness standard. 

In \S\ref{sec:param_compare}, we argue that the parameter sets making up each of these models are discrepant at the $\sim$6$\sigma$ level.  Curiously, the amplitude parameters $A_s$ differ by less than 0.1$\sigma$, so the differences noted above are being absorbed by a combination of all the \lcdm\ parameters.  Even so, the most discrepant parameter, $\Omega_c h^2$, differs by only 1.7$\sigma$ (when all other parameters are marginalized over).   The announcement accompanying the first \planck\ data release emphasized that the universe is older and more massive than previously believed.  This change is largely driven by the higher $\Omega_c h^2$ that \planck\ infers, though differences in the inferred Hubble constant \citep{bennett/etal:2014} may also contribute.

In this paper we examine differences between the \planck\ and \wmap\ data in detail.   In \S\ref{sec:dipole_compare} we study large angular scale differences in the delivered, single-frequency sky maps.  In \S\ref{sec:fg_cln_compare} we correct for residual foreground differences and again compare single-frequency map differences.  In \S\ref{sec:multi_map_compare} we use a modified internal linear combination technique to examine multi-frequency cosmic microwave background (CMB) maps.  The \planck\ multi-frequency SMICA map is compared with the \wmap\ internal linear combination (ILC) map as well.

In \S\ref{sec:spec_compare} we compare the angular power spectra from the two experiments and address calibration issues.  We start in \S\ref{sec:spec_models} with a comparison of the \lcdm\ model fits (without regard to the inferred parameter values) as a means to enhance the sensitivity of the comparison.  We argue that the shape of the power spectrum difference likely arises from errors in the transfer function of one or both experiments.  Unfortunately we lack access to the requisite data needed to fully assess the \planck\ transfer function, but in \S\ref{sec:wmap_beam} we re-examine the \wmap\ transfer function, specifically the \wmap\ main beam profiles and solid angle estimates in \S\ref{sec:re_beam}, and the \wmap\ far sidelobe response and and its effect on dipole calibration in \S\ref{sec:re_sidelobe}.  These re-analyses are aimed at establishing the largest potential uncertainty in the \wmap\ measurements of the CMB power spectrum.

In \S\ref{sec:param_compare} we compare the cosmological parameters inferred from the \planck\ and \wmap\ data.  In addition to individual parameter comparisons, we study the consistency of the full 6-parameter \lcdm\ model as a whole, taking into account the effects required to produce an accurate covariance matrix of parameter differences.  The 6 \lcdm\ parameters are the physical baryon density, $\Omega_b h^2$; the physical cold dark matter density, $\Omega_c h^2$; the amplitude of scalar fluctuations at $k_0 = 0.05$ Mpc$^{-1}$, $A_{s,0.05}$; the scalar fluctuation power spectral index, $n_s$; the  Hubble constant, $H_0$, in km s$^{-1}$ Mpc$^{-1}$; and the reionization optical depth, $\tau$.  Our conclusions are presented in \S\ref{sec:conclusions}.

\section{Sky map comparisons}\label{sec:map_compare}

We compare released versions of co-added single-frequency sky maps from \planck\ and \wmap, especially those derived from observations taken using similar pass bands.   We also compare results taken from linear combinations of multi-frequency maps.  All comparisons are performed with maps smoothed to a common resolution of $2^\circ$.  Examination of maps on these scales is useful for visualizing large-scale effects such as far-sidelobe pickup.

\begin{figure*}[ht]
\begin{center}
\includegraphics[height=3.0in]{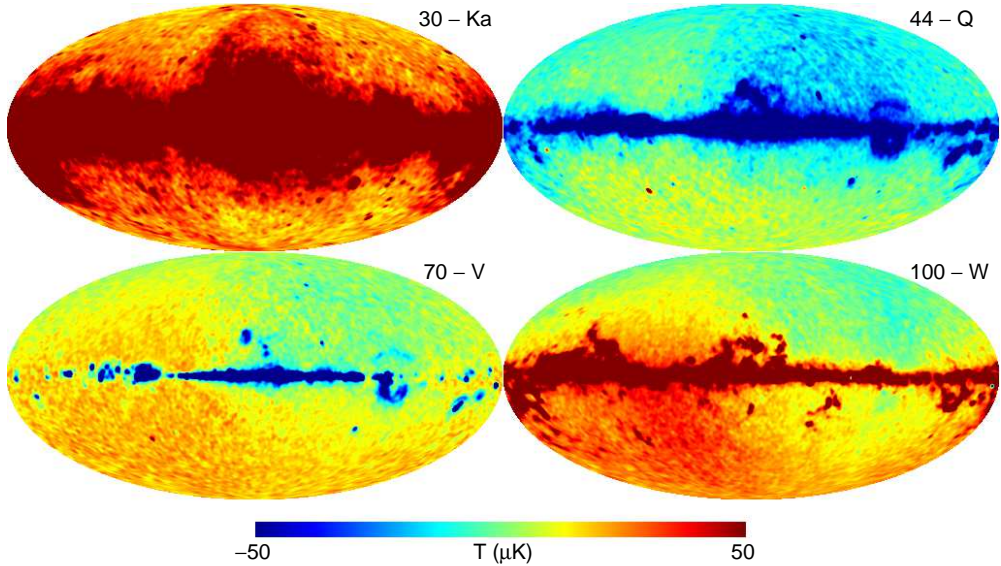}
\end{center}
\caption{Differences between nearest-frequency \planck\ and \wmap\ sky maps, smoothed to 2$^\circ$.  The \planck\ bands are designated by their nominal frequencies (30, 44, 70 and 100 GHz), while \wmap\ bands are designated by Ka, Q, V and W, corresponding to nominal frequencies of 33, 41, 61 and 94 GHz, respectively.  Differences are in thermodynamic units, on a $\pm$50 $\mu$K scale.  Since the \planck\ and \wmap\ pass bands are not identical, Galactic foreground residuals are present; this is largest at the lower frequencies: the 30$-$Ka difference is saturated in the plane on this scale.  The differences are always shown as \planck\ minus \wmap\, regardless of frequency ordering; this produces the expected negative Galactic residual in 44$-$Q and 70$-$V.  Additionally, the \planck\ 100 GHz band encompasses emission from the 115 GHz CO $J=1-0$ transition, which is seen in the Galactic plane in the 100$-$W difference.  Although far-sidelobe (FSL) pickup of dipole and Galactic signals has been removed from the \wmap\ maps, it is still present in the currently available \planck\ maps at these frequencies.  Of the three LFI bands, FSL pickup is strongest at 30 GHz.  A residual dipole is evident in the 44$-$Q, 70$-$V and 100$-$W maps, indicative of calibration differences between the two experiments.  A dipole residual similar to that seen in 44$-$Q and 70$-$V is also present in 30$-$Ka, but is obscured by the Galactic foreground residual, and by the presence of \Planck\ 30 GHz far-sidelobe pickup of the CMB dipole.  A deeper investigation of underlying differences between the two experiments at spatial scales $l>1$ necessitates the removal of dipole and foreground residuals, and also ideally the \Planck\ FSL contribution. \label{fig:raw_nearfreq}}
\end{figure*}

\subsection{Published single-frequency maps}\label{sec:dipole_compare}

The three \Planck\ LFI bands and the 100~GHz HFI band have similar frequency coverage to the \wmap\ Ka, Q, V and W bands.  We form \planck$-$\wmap\ difference maps from the pairs $30-$Ka, $44-$Q, $70-$V, and $100-$W in thermodynamic temperature units. If the absolute calibration of the \planck\ and \wmap\ maps agree to within a percent or so, the $l>1$ CMB anisotropy signal should visually cancel in the difference maps.  However, because the CMB dipole ($l=1$) signal is relatively bright, even small calibration differences will be expected to produce residual dipoles in the difference maps.  Further, since the \planck\ and \wmap\ band pairs have somewhat different pass bands, we expect to see imperfect foreground signal cancellation in the difference maps.  Figure~\ref{fig:raw_nearfreq} shows sky map differences formed from the nominal 15.5 month \planck\ DR1 release and the final nine-year \wmap\ release.

The [44,Q], [70,V] and [100,W] map differences in Figure~\ref{fig:raw_nearfreq} show clear dipole residuals.  (The [30,Ka] difference map is dominated by Galactic signals and \planck\ far-sidelobe pickup, so it requires further processing to uncover the dipole residual.)  These dipoles are well aligned with the CMB dipole direction, and show a systematic sign: all are positive in the south and negative in the north.  This sign is consistent with the \wmap\ sky maps having a brighter temperature scale than \Planck's.  We use the seven-year \wmap\ value of the CMB dipole \citep{jarosik/etal:2011} to estimate the fractional magnitude of the residual in each of these maps.  We find the residual dipoles in the [44,Q], [70,V] and [100,W] differences are 0.3\%, 0.3\% and 0.5\% of the CMB dipole, respectively.  These values are within the quoted uncertainties for both the \planck\ and \wmap\ data \citep{planck/03:2013, planck/08:2013, bennett/etal:2013}.   Note that these differences are insufficient by at least a factor of 2 to explain the power spectrum differences at higher multipoles (see \S\ref{sec:spec_models}).

\subsection{Foreground-reduced single frequency maps}\label{sec:fg_cln_compare}

Contributions from Galactic foregrounds must be removed (`cleaned') from the single-frequency difference maps before any residuals not associated with true sky signals may be seen.  In this section, that foreground removal is accomplished using spatial templates that trace known Galactic emission, fit as a linear combination to each difference map.   The templates used for this purpose include some that are completely external to either the \planck\ or \wmap\ data, as well as pairwise differences between \wmap\ bands.  The cleaned difference maps are largely independent of the exact templates used for cleaning; this is especially true of the cleaned 70$-$V and 100$-$W maps, where Galactic foregrounds are near their minimum.

\begin{figure*}[htb]
\begin{center}
\includegraphics[height=5in]{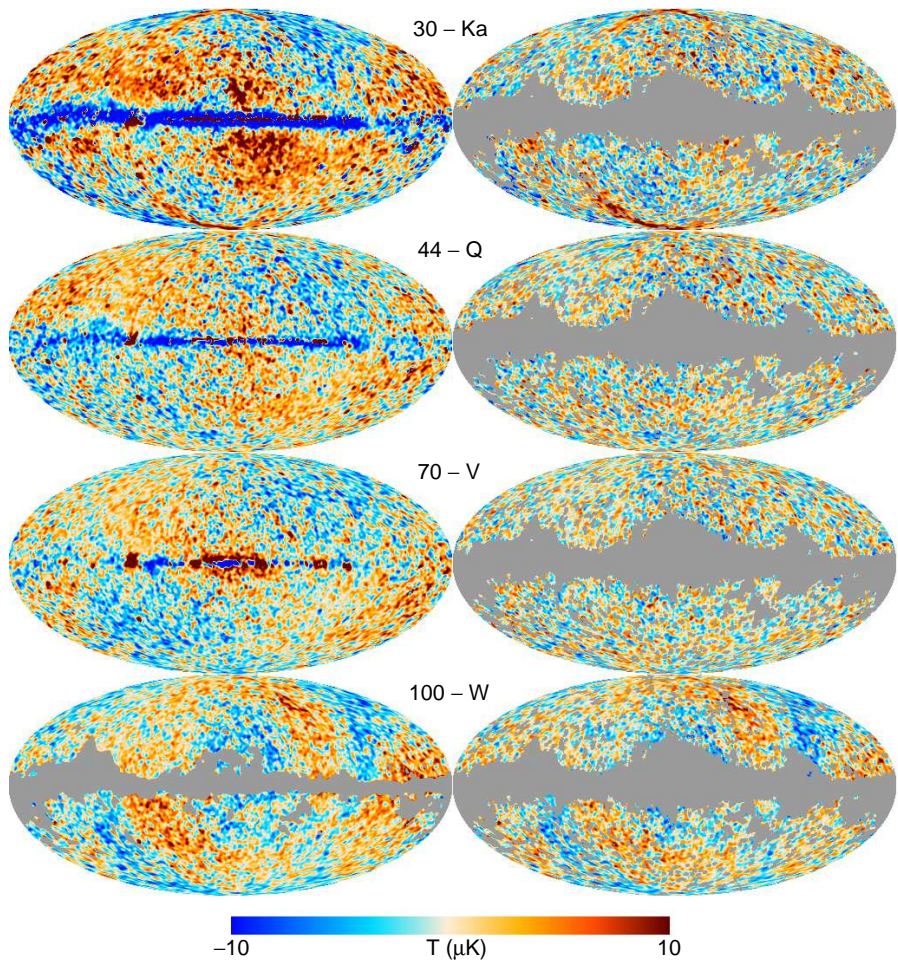}
\end{center}
\caption{Template-cleaned \planck$-$\wmap\ difference maps; the color scale is $\pm 10\, \mu$K.  The maps have been smoothed to 2$^\circ$ resolution.  Each row corresponds to a single-frequency difference, from top to bottom: 30$-$Ka, 44$-$Q, 70$-$V, and 100$-$W. \textsl{Left}: residuals after removal of foreground templates and a dipole component.  The template fits were performed over the entire sky for the LFI$-$\wmap\ differences: thus as expected, foreground removal is imperfect near the Galactic plane.  Regions with strong CO emission (grey) were excluded from the 100$-$W template fit.  \textsl{Right}: residuals after removal of foreground templates and both a dipole and quadrupole component.  Grey areas indicate regions masked from the fit, which was kept the same for all maps in this column.  Inclusion of quadrupole terms in the template fits removes much of the residual signal from the LFI$-$\wmap\ differences.  No far sidelobe correction has been applied to \planck\ data; far sidelobe pickup of the CMB dipole is removed to first order by inclusion of a dipole fitting component.  Far sidelobe Galactic pickup remains; this is most evident in 30$-$Ka, where Galactic far sidelobe pickup produces a $\sim$85$^\circ$ ``ring'' about the Galactic center.  \label{fig:cleaned_nearfreq}}
\end{figure*}

The left column of Figure~\ref{fig:cleaned_nearfreq} shows the \Planck$-$\wmap\ single-frequency difference maps after removal of foreground templates and a dipole component.  Each LFI$-$\wmap\ difference is cleaned using a linear combination of 3 foreground templates: two in common and one specific to a given frequency pair.  The first of the two common templates is the FDS dust emission model~8 \citep{finkbeiner/davis/schlegel:1999} evaluated at 94 GHz which, for our purposes, serves as either a thermal or spinning dust template.  The second common template is a surrogate for free-free emission: we employ a modified version of the \citet{finkbeiner:2003} composite H$\alpha$ template which has been corrected for Galactic extinction \citep{bennett/etal:2013}.  A third template is formed from the difference between bracketing \wmap\ frequencies: for example, K$-$Q was used as a template for the 30$-$Ka pair.  This choice permits a realistic near-frequency foreground template at the expense of including \wmap\ instrument noise.  For 100$-$W, only the two common foreground templates were used, but high CO emission near the Galactic plane in the \planck\ 100~GHz map necessitated masking this region from the template fit.

The LFI$-$\wmap\ differences in the left column of Figure~\ref{fig:cleaned_nearfreq} exhibit a clear quadrupolar signature. This originates in part from the treatment of the kinematic quadrupole in the delivered \planck\ and \wmap\ map products.   Both HFI and \wmap\ data retain the kinematic quadrupole in the map products \citep{planck/08:2013, jarosik/etal:2007} whereas LFI intended to fully remove it \citep{planck/02:2013}. However, the \planck\ Collaboration later noted that a code bug had unintentionally caused the retention of half of the kinematic quadrupole \citep[Appendix A.1]{planck/05:2013} in the 2013 DR1 LFI maps, leaving a residual quadrupole with a peak-to-trough amplitude of $\sim$2 $\mu$K.  We have additionally fit for and removed a quadrupole from the difference maps and the result is shown in the right column of Figure~\ref{fig:cleaned_nearfreq}.  This removal leaves a recognizable \planck\ FSL signature in the 30$-$Ka difference map, but the majority of structure in the 44$-$Q and 70$-$V difference maps appears to have been removed.  The quadrupole fit to the 30$-$Ka map has a peak-to-trough amplitude near 8 $\mu$K and is most likely influenced by residual FSL and foregrounds.  The quadrupoles fit to the 44$-$Q and 70$-$V maps have similar peak-to-trough amplitudes of roughly 4 $\mu$K and are fairly well aligned with the residual kinematic quadrupole, but have amplitudes twice that expected based on the documented treatment of the kinematic quadrupole in LFI and WMAP data processing.  It is possible, however, that the residual quadrupole we observe is associated with a systematic which just happens to align with the kinematic quadrupole.  A remaining known systematic is the LFI FSL pickup which was not removed in the delivered sky maps; we next attempt to remove this contribution based on published LFI estimates. 

\begin{figure*}[ht]
\begin{center}
\includegraphics[height=5in]{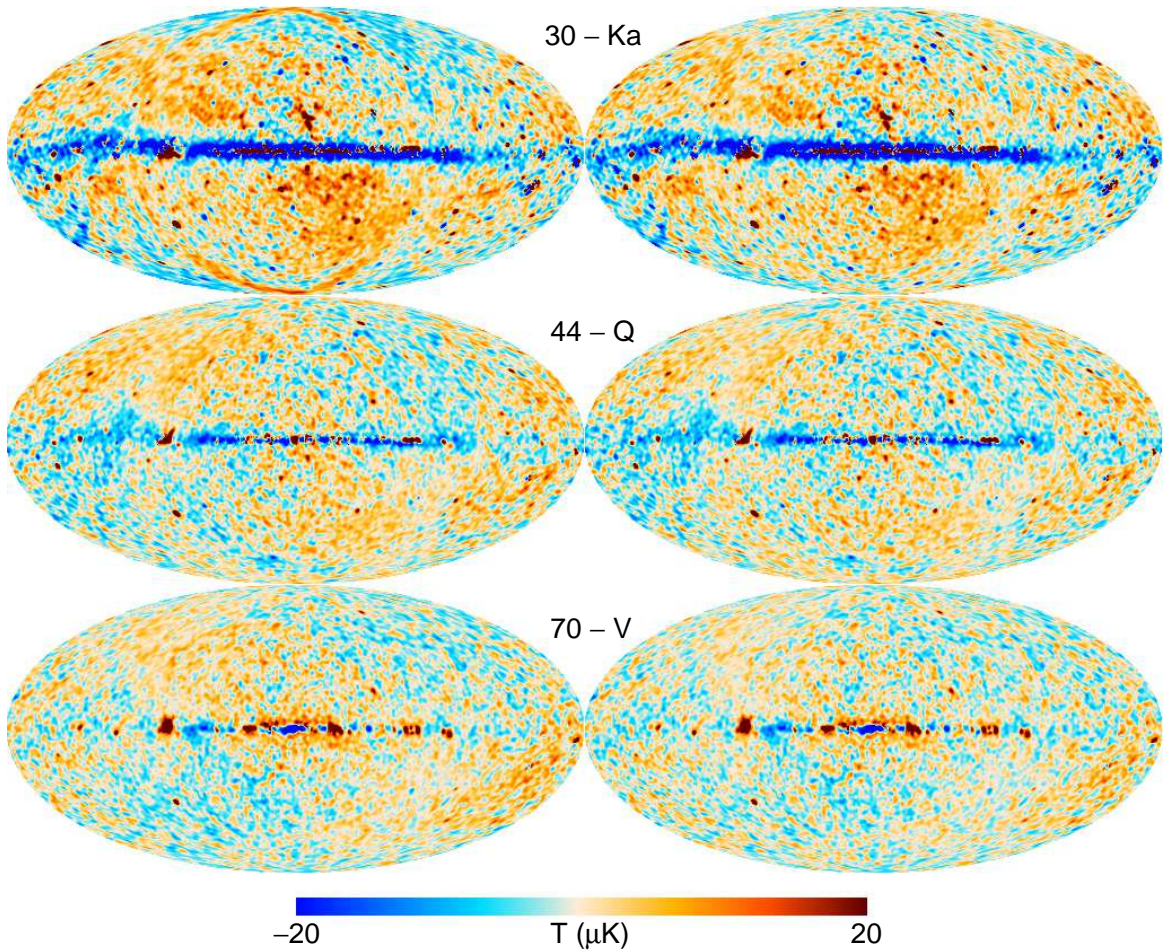}
\end{center}
\caption{Combined \planck\ (Survey~1$+$Survey~2)$-$\wmap\ sky maps; the color scale is $\pm$20 $\mu$K.  The maps have been smoothed to $2^\circ$ resolution.  Each row corresponds to a single-frequency difference: from top to bottom these are 30$-$Ka, 44$-$Q, and 70$-$V.  The left column shows differences after removal of Galactic emission, dipole and monopole components.  The right column is identical to the left, with the exception that an estimate of FSL emission and a half-strength kinematic quadrupole component have been removed.  The \planck\ FSL contribution was obtained using digitized versions of the estimates published in \citet{planck/03:2013}.  The quadrupolar structure seen in Figure~\ref{fig:cleaned_nearfreq} is present in these difference pairs as well.  \label{fig:s1s2_fslcorr_cleaned_nearfreq}}
\end{figure*}

Figure~\ref{fig:s1s2_fslcorr_cleaned_nearfreq} illustrates results similar to Figure~\ref{fig:cleaned_nearfreq} but uses only the \planck\ Survey 1 and Survey 2 LFI maps, rather than the nominal 15.5 month mission maps.  The reason for using only Survey 1 and 2 maps is two-fold: 1) we can remove digitally scanned versions of LFI far-sidelobe estimates for Survey 1 and 2 in order to estimate the effect of FSL removal, and 2) estimates of LFI systematics are not currently provided for Survey~3 \citep{planck/03:2013}.  The Galactic foreground templates and coefficients used to clean the (Survey 1$+$Survey 2)$-$\wmap\ data are identical to those used to clean the nominal 15.5 month \planck$-$\wmap\ difference maps.   As seen in the right-hand column of Figure~\ref{fig:s1s2_fslcorr_cleaned_nearfreq}, removal of the digitized FSL pickup does not completely remove high-latitude, large-scale structure in the LFI$-$\wmap\ differences.  We conclude that unaccounted-for systematics remain, but at this stage cannot attribute them unambiguously to a particular instrument.

\begin{figure}[ht]
\begin{center}
\includegraphics[width=0.45\textwidth]{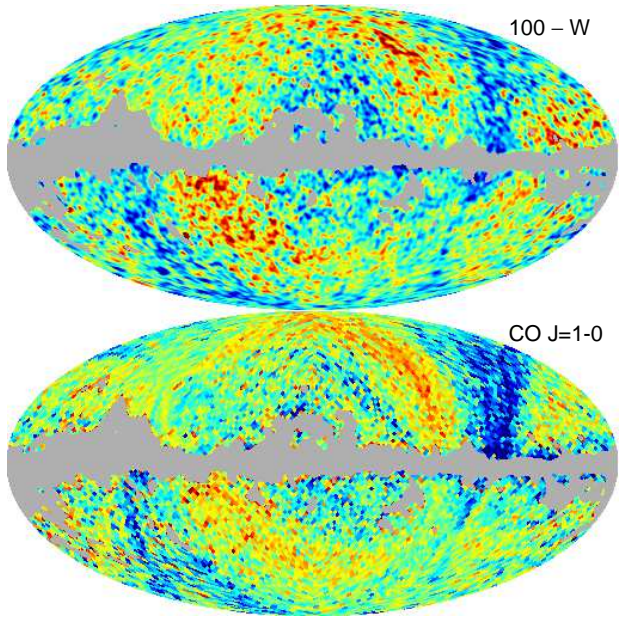}
\end{center}
\caption{\textsl{Top}: \planck$-$\wmap, for the 100$-$W pair, after template subtraction of foreground emission and  smoothing to $2^\circ$ resolution; the color scale is $\pm$10 $\mu$K.  The grey region masks strong CO emission present in the \planck\ 100 GHz map.  There is clear residual structure in excess of the instrument noise.  \textsl{Bottom}:  The \planck\ Type 2 CO $J=1-0$ map, degraded to $N_{\rm side}= 32$ ($1.8^\circ$ pixels).  As in the top image, the grey region masks most of the known CO emission.  The remaining large-scale features are unrelated to CO emission and may be considered residual systematics. A logarithmic scale is used to emphasize these features.  The similarity of the structure in these two map suggests that the structure in the difference map originates in the \planck\ 100 GHz data, rather than the W-band map, since the CO map is made entirely with \planck\ data.  \label{fig:CO_emission}}
\end{figure}

The 100$-$W difference map has a different residual signature than that seen in the LFI$-$\wmap\ maps.  The residuals are shown both in Figure~\ref{fig:cleaned_nearfreq} and in the top panel of Figure~\ref{fig:CO_emission}.  Figure~\ref{fig:cleaned_nearfreq} demonstrates that visible structure in 100$-$W is little affected by removal of a quadrupole component.  While searching for possible explanations of the structure in this map, we coincidentally examined the \planck-estimated maps of CO emission, in particular the Type 2 and 3 CO maps, which are generated solely from \planck\ data \citep{planck/13:2013}.  The bottom panel of Figure~\ref{fig:CO_emission} shows high latitude, low-level residuals not associated with CO emission present in the Type 2 CO map, which is produced from \Planck\ 70, 100, 143 and 353 GHz data. (The Type 3 CO map has similar structure, but is not shown.)  The similarity of the top and bottom images gives us reasonable confidence that most of the structure in the 100$-$W map arises from the 100 GHz map.  We discuss this further in the next section.

\subsection{Multi-frequency difference maps}\label{sec:multi_map_compare}

\subsubsection{Extended ILCs}

In this section we consider another approach to foreground subtraction.  As originally implemented, the Internal Linear Combination (ILC) method \citep{bennett/etal:2003} was designed to produce a foreground-cleaned CMB map from a linear combination of multi-frequency maps, each in thermodynamic temperature units.  The single-frequency map weights were chosen to minimize the variance of the combined map (within a large multi-pixel region) subject to the constraint that the weights sum to 1.0, thus preserving the CMB signal.

In this section we extend the ILC algorithm to form what may be regarded as the difference between two ILC's.  The algorithm, described 
in greater detail in Appendix~\ref{sec:eilc_app}, still uses a minimum variance criteria to minimize foreground signals, but it breaks the input maps into two groups: the weights for the first group are constrained to sum to $+$1, the second group to $-1$.  Ideally, the map formed in this way would completely cancel all CMB and foreground signals, yielding a null map. Systematic effects that are common between the two groups of maps would also tend to cancel, leaving behind signals that are distinct between the two sets.

We form extended ILC (eILC) maps from three groupings of data: LFI$-$\wmap, HFI$-$\wmap, and HFI$-$LFI.  We use all the available LFI and \wmap\ frequencies, but only the 100, 143 and 217 GHz maps from HFI.  We further limit the weight computation to use only pixels outside of the \wmap9 KQ75 mask \citep{bennett/etal:2013}, which avoids strong foreground emission in the Galactic plane, in particular strong CO contamination in the HFI 100 and 217 GHz maps.  In addition, we add back the half-strength kinematic quadrupole removed in the LFI processing \citep{planck/05:2013} to ensure quadrupole consistency among the groupings.

\begin{figure*}[ht]
\begin{center}
\includegraphics[width=6.5in]{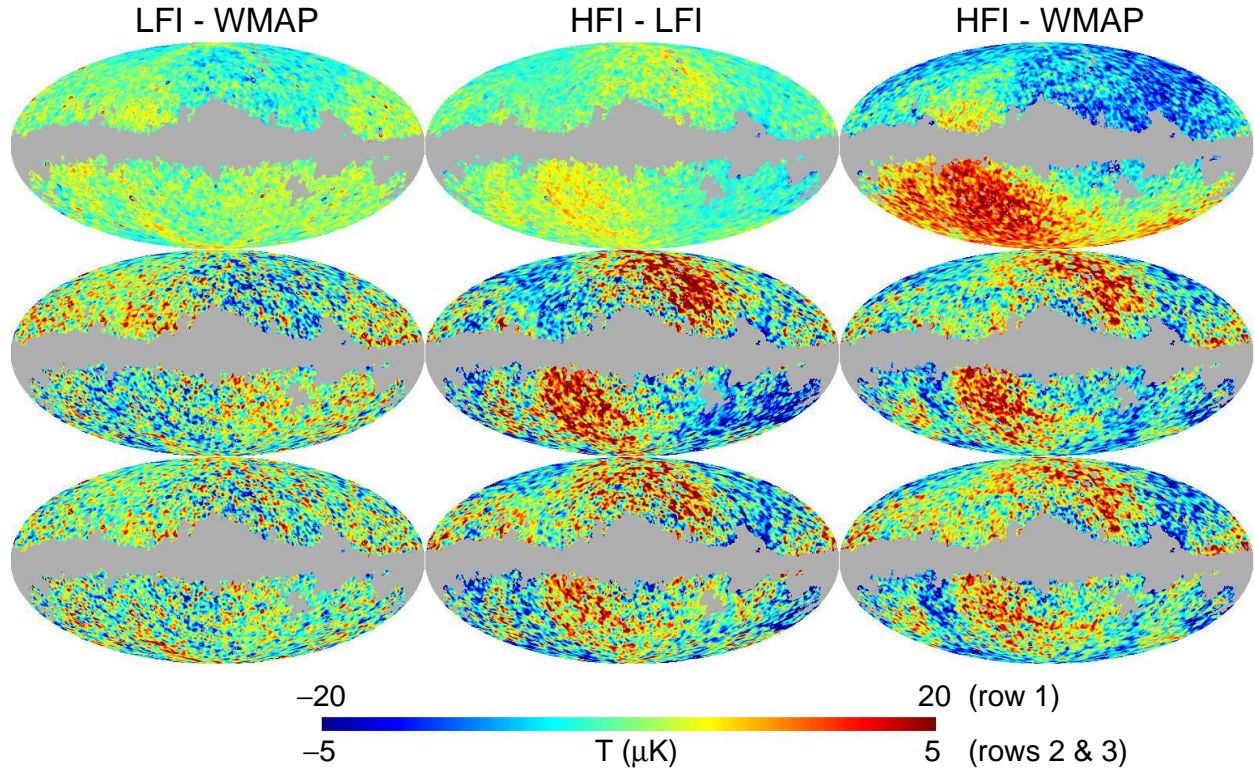}
\end{center}
\caption{Extended ILC maps computed using maps smoothed to 2$^\circ$ resolution and applying a KQ75 mask (grey regions).  In the absence of systematics, these maps should resemble noise, but this is clearly not the case.  The left column is LFI$-$\wmap, the middle is HFI$-$LFI, and the right is HFI$-$\wmap.  Only 100, 143, and 217 GHz maps are used for the HFI calculations.  The top row has no dipole component removed and uses a color scale of $\pm$20 $\mu$K.  The middle row has a dipole component removed and uses a color scale of $\pm$5 $\mu$K.  The bottom row has a dipole and quadrupole removed and uses a color scale of $\pm$5 $\mu$K.  The weights are given in Table~\ref{tab:weights2_5uK_no353}.  The \planck-only HFI$-$LFI combination indicates that systematic structure seen in the \planck$-$\wmap\ combinations arises primarily from the \planck\ maps.  \label{fig:weights2_5uK_no353}}
\end{figure*}

The eILC maps from the groupings LFI$-$\wmap, HFI$-$\wmap, and HFI$-$LFI are shown in the first row of Figure~\ref{fig:weights2_5uK_no353}.  The weights used to form these maps are given in the first column of Table~\ref{tab:weights2_5uK_no353}, under the columns labeled $a$.  Evidently a dipole component must be removed from these maps in order to view finer details.  To facilitate this, we add dipole components to the eILC computation, as described in Appendix~\ref{sec:eilc_app}, and recompute the weights (the columns labeled $b$ in Table~\ref{tab:weights2_5uK_no353}).  The results are shown in the middle row of Figure~\ref{fig:weights2_5uK_no353}.  The last row in Figure~\ref{fig:weights2_5uK_no353} similarly shows the result of including both dipole and quadrupole components in the eILC computation (the weights are given in the columns labeled $c$ in Table~\ref{tab:weights2_5uK_no353}).

The sky maps in the middle row of Figure~\ref{fig:weights2_5uK_no353} are similar in appearance to their counterparts constructed from template-cleaned, single-frequency difference maps, as discussed in \S\ref{sec:fg_cln_compare}.   The LFI$-$\wmap\ eILC exhibits a quadrupole structure similar to that noted in the 30$-$Ka, 44$-$Q and 70$-$V differences.  The HFI$-$\wmap\ eILC shows a similar ring-like structure as seen in 100$-$W.  The \planck-only HFI$-$LFI eILC combination appears to contain both the broad ring-like structure and the inverse of the quadrupole signature noted in the \planck$-$\wmap\ combinations.  Taken together, these features argue for the presence of  residual systematic effects in the \planck\ maps, at the level of $\sim$5 $\mu$K.

The presence of low-level residuals in the \planck\ DR1 maps is not unexpected.  Both the LFI and HFI papers note incomplete removal of far-sidelobe pickup from the delivered maps \citep{planck/03:2013, planck/14:2013}.  Other systematics in the LFI data are discussed in \citet{planck/03:2013}, which states that the main effects are due to ``stray-light pick-up from far sidelobes and imperfect photometric calibration''.   We cannot definitively attribute residuals to a single cause.  It is tempting to associate the broad ring-like structure in the 100$-$W difference map with \planck\ FSL pickup, since there is a known ring from primary spillover located roughly 85$^\circ$ from the Galactic center.  However, it seems unlikely that the eILC residual is due to direct FSL pickup because the feature is not sharp.  It appears more likely to us to be an indirect effect of FSL pickup on calibration, such as that illustrated in Figure 11 of \citet{planck/05:2013}.  Alternatively, these features may arise from time-dependent calibration residuals due to ADC nonlinearity or thermal effects (see, e.g., Figures 15 and 28 of \citet{planck/03:2013}).   

\capstartfalse
\begin{deluxetable*}{crrrrrrrrr}
\tablecaption{Extended ILC Coefficients\label{tab:weights2_5uK_no353}}
\tablewidth{0pt}
\tablehead{\colhead{Band/} & \multicolumn{3}{c}{HFI$-$LFI} & \multicolumn{3}{c}{LFI$-$WMAP} & \multicolumn{3}{c}{HFI$-$WMAP} \\
           \colhead{Component} & \colhead{\tablenotemark{a}} & \colhead{\tablenotemark{b}} & \colhead{\tablenotemark{c}} & 
	   \colhead{ \tablenotemark{a}} & \colhead{\tablenotemark{b}} & \colhead{\tablenotemark{c}} &
	   \colhead{ \tablenotemark{a}} & \colhead{\tablenotemark{b}} & \colhead{\tablenotemark{c}}}
\startdata
K & --\hspace{1em} & --\hspace{1em} & --\hspace{1em} & $-0.237$ & $-0.002$ & $-0.010$ & $-0.239$ & $0.039$ & $0.034$ \\
Ka & --\hspace{1em} & --\hspace{1em} & --\hspace{1em} & $-0.133$ & $-0.176$ & $-0.187$ & $1.485$ & $0.060$ & $0.067$ \\
Q & --\hspace{1em} & --\hspace{1em} & --\hspace{1em} & $-0.383$ & $-0.315$ & $-0.299$ & $-1.313$ & $-0.172$ & $-0.184$ \\
V & --\hspace{1em} & --\hspace{1em} & --\hspace{1em} & $-0.145$ & $-0.332$ & $-0.315$ & $-0.808$ & $-0.482$ & $-0.488$ \\
W & --\hspace{1em} & --\hspace{1em} & --\hspace{1em} & $-0.102$ & $-0.174$ & $-0.189$ & $-0.124$ & $-0.445$ & $-0.429$ \\
\hline
30 &  $0.202$ &  $0.122$ &  $0.107$ & $0.669$ & $0.149$ & $0.175$ & --\hspace{1em} & --\hspace{1em} & --\hspace{1em} \\
44 & $-0.552$ & $-0.276$ & $-0.285$ & $-0.101$ & $0.306$ & $0.304$ & --\hspace{1em} & --\hspace{1em} & --\hspace{1em} \\
70 & $-0.650$ & $-0.846$ & $-0.823$ & $0.432$ & $0.545$ & $0.521$ & --\hspace{1em} & --\hspace{1em} & --\hspace{1em} \\
\hline
100 & $0.448$ & $0.452$ & $0.389$ & --\hspace{1em} & --\hspace{1em} & --\hspace{1em} & $0.232$ & $0.402$ & $0.360$ \\
143 & $0.812$ & $0.784$ & $0.854$ & --\hspace{1em} & --\hspace{1em} & --\hspace{1em} & $1.055$ & $0.821$ & $0.868$ \\
217 & $-0.261$ & $-0.237$ & $-0.243$ & --\hspace{1em} & --\hspace{1em} & --\hspace{1em} & $-0.287$ & $-0.223$ & $-0.228$ \\
\hline
$d_x$ & --\hspace{1em} & $-0.225$ & $-0.229$ & --\hspace{1em} & $-0.675$ & $-0.657$ & --\hspace{1em} & $-1.130$ & $-1.124$ \\
$d_y$ & --\hspace{1em} & $-2.138$ & $-2.176$ & --\hspace{1em} & $-6.401$ & $-6.242$ & --\hspace{1em} & $-10.735$ & $-10.678$ \\
$d_z$ & --\hspace{1em} &  $2.409$ &  $2.452$ & --\hspace{1em} & $7.223 $ & $7.035$  & --\hspace{1em} &  $12.098$ & $12.034$ \\
\hline
$\tilde{a}_{2,-2}$ & --\hspace{1em} & --\hspace{1em} & $0.077$ & --\hspace{1em} & --\hspace{1em} & $0.200$ & --\hspace{1em}  & --\hspace{1em} & $1.279$ \\
$\tilde{a}_{2,-1}$ & --\hspace{1em} & --\hspace{1em} & $4.034$ & --\hspace{1em} & --\hspace{1em} & $-1.764$ & --\hspace{1em} & --\hspace{1em} & $2.102$ \\
$\tilde{a}_{2,0}$ & --\hspace{1em} & --\hspace{1em} & $-1.838$ & --\hspace{1em} & --\hspace{1em} &  $1.890$ & --\hspace{1em} & --\hspace{1em} & $-0.691$ \\
$\tilde{a}_{2,1}$ & --\hspace{1em} & --\hspace{1em} & $-1.195$ & --\hspace{1em} & --\hspace{1em} & $-0.139$ & --\hspace{1em} & --\hspace{1em} & $-0.448$ \\
$\tilde{a}_{2,2}$ & --\hspace{1em} & --\hspace{1em} & $ 1.136$ & --\hspace{1em} & --\hspace{1em} & $-0.851$ & --\hspace{1em} & --\hspace{1em} & $0.682$ 
\enddata
\tablecomments{The weights in the first group (HFI or LFI) sum to $+1$, and in the second group (LFI or WMAP) to $-1$. The weights for the band maps are dimensionless, the weights for the multipole components have units of $\mu$K.  Harmonic conventions are defined in Appendix~\ref{sec:conventions}.}
\tablenotetext{a}{No multipole components were included in the weight computation.}
\tablenotetext{b}{Three dipole components were additionally included in the weight computation.}
\tablenotetext{c}{Five quadrupole components were additionally included in the weight computation.}
\end{deluxetable*}
\capstarttrue

\subsubsection{\Planck\ SMICA and \wmap\ ILC}

\begin{figure}[ht]
\begin{center}
\includegraphics[width=0.45\textwidth]{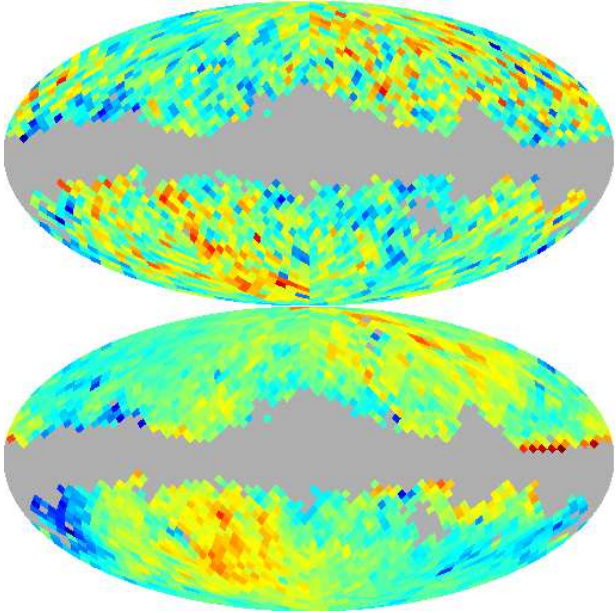}
\end{center}
\caption{\textsl{Top}: the SMICA$-$ILC map at HEALPix $N_{\rm side}=16$ ($3.6^\circ$ pixels); the color scale is $\pm$20 $\mu$K. 
\textsl{Bottom}: the ``SMICA artifact map'' constructed from \planck\ data as described in the text.  This artifact map produces similar large-scale structure to that seen in the SMICA$-$ILC difference map.  \label{fig:smica_minus_ilc}}
\end{figure}

We next compare CMB maps produced by the \planck\ and \wmap\ teams, respectively. The \wmap\ product is referred to simply as the ILC map \citep{bennett/etal:2013}.  \citet{planck/12:2013} discuss four different methods for producing foreground-cleaned CMB maps.  Of these, we adopt the SMICA map as representative of this class of \planck\ DR1 products, based on the \Planck\ Collaboration's recommendation \citep{planck/12:2013}.  We obtain the SMICA map and the \wmap\ nine-year ILC map at the same HEALPix resolution, smooth both to a common $2^\circ$ FWHM resolution, and difference them.  We further remove a dipole outside of a mask, and degrade to HEALPix $N_{\rm side}=16$ ($3.6^\circ$ pixels) to reduce the noise.  This difference map is shown in the top image of Figure~\ref{fig:smica_minus_ilc}.

Large-scale structure in the SMICA$-$ILC difference map has been noted previously, e.g. \citet{kovacs/etal:2013}, but we are unaware of any explanation of it in the literature.  Based on our findings in previous sections, we conjecture that the dominant large-scale features arise from residual systematics in the \planck\ maps.  We can test this hypothesis by forming maps of estimated systematics at each \planck\ frequency and combining them using the published SMICA weights to form an ``artifact SMICA''.  If the hypothesis is plausible, then the ``artifact SMICA'' will be similar to the SMICA$-$ILC map itself.

To estimate the residual systematics at each \planck\ frequency, we start with a 100$-$70 GHz difference map, and remove residual foregrounds using a linear combination of the FDS dust model and the Finkbeiner H$\alpha$ template, as in \S\ref{sec:fg_cln_compare}.  The resultant foreground-cleaned map looks very much like the HFI$-$LFI eILC in Figure~\ref{fig:weights2_5uK_no353}.  A quadrupole component is additionally removed to form the 100~GHz template.  This template resembles the cleaned 100$-$W difference map, but is free of any \wmap\ data.  Next, candidate templates for 143, 217 and 353 GHz are constructed by differencing each of these band maps against the 100 GHz map, cleaning foregrounds with the above dust and free-free templates, and then adding back the 100 GHz artifact template.  For 545 and 857 GHz, we adopt the far-sidelobe maps provided in the \planck\ DR1 release, but with the zodiacal contribution removed.  For 30, 44 and 70 GHz, we adopt the small excess quadrupole in conjunction with the digitized LFI FSL pickup maps from \S\ref{sec:fg_cln_compare}.

Once we have candidate templates for each \planck\ frequency, we combine them using SMICA weights applicable to $l < 50$.  We digitally scan these values from Figure~D.1 of \citet{planck/12:2013}.  Weights are specified in units such that a SMICA map in thermodynamic temperature units is produced from a linear combination of maps in Rayleigh-Jeans units.  The weights we use are listed in Table~\ref{tab:smica_wts}.  The frequencies with the most weight are 100, 143, 217 and 353 GHz.  Our estimated SMICA artifact map is shown in the bottom panel of Figure~\ref{fig:smica_minus_ilc}, and may be directly compared to the SMICA$-$ILC residual map above it.  We conclude that this model for the SMICA$-$ILC residual is plausible; but other plausible models may exist.

\capstartfalse
\begin{deluxetable}{cr}
\tablecaption{Low multipole SMICA weights\tablenotemark{a}\label{tab:smica_wts}}
\tablewidth{0pt}
\tablehead{\colhead{Band} & \colhead{Weight}  \\
        \colhead{[GHz]}  & \colhead{[$\mu$K$^{\rm CMB}$/$\mu$K$^{\rm RJ}$]}}
\startdata
30  &  $+$0.009708 \\
44  &  $-$0.1029\phm{00} \\
70  &  $-$0.1145\phm{00} \\
100  &  $-$0.3281\phm{00} \\
143  &  $+$1.95\phm{0000}  \\
217  &  $+$1.303\phm{000} \\
353  &  $-$2.242\phm{000} \\
545  &  $+$0.8252\phm{00} \\
857  &  $-$0.13\phm{0000} 
\enddata
\tablenotetext{a}{These weights have been digitally scanned from Figure~D.1 of \citet{planck/12:2013}.  They are applicable only to $l < 50$ and subject to digitization error.} 
\end{deluxetable}
\capstarttrue

\section{Power spectrum comparison}\label{sec:spec_compare}

The \wmap\ and \planck\ temperature measurements can be compared in a variety of ways over a range of angular scales.   The results appear to present a complex and sometimes conflicting story: at large to medium angular scales, the \planck\ data calibration produces a fainter sky than \wmap, while at the smallest angular scales within the 30-70 GHz bands, \planck\ returns a slightly brighter surface temperature for Jupiter than does \wmap, \citep{planck/05:2013}, though they are within the 0.5\% to 1\% uncertainties.

On the largest scale, the dipole residual in the \wmap\ $-$ \planck\ difference map is $<$ 0.5\% of the CMB dipole (\S\ref{sec:dipole_compare}).  Other systematic low-$l$ structure, on the scale of a few $\mu$K, is present in both the LFI and HFI \planck\ maps (\S\ref{sec:map_compare}), but these effects are confined to roughly $l \le 10$, and they contribute only a few $\mu {\rm K}^2$ to the angular power spectrum.  This is not a major source of discrepancy with \wmap.  At intermediate angular scales, the discrepancy is more significant, as noted in \citet{planck/31:2013}; see also Figure~\ref{fig:planck_wmap_powspec_ratio} above.

\subsection{Power spectrum model fits}\label{sec:spec_models}

The largest difference between \wmap\ and \planck\ lies in the temperature power spectrum at $\ell \gtrsim 100$.  To quantify the significance of this difference in a way that captures the full error estimates provided by each team, we evaluate the quantity $D_{220} \equiv l(l+1)C^{\rm TT}_l/(2\pi) |_{l=220}$, the model spectrum amplitude near the first acoustic peak, using the Markov chains released by each team.  We choose to compare the model amplitude (assuming flat \lcdm) because it effectively smooths the data over a range of $l$, substantially reducing the uncertainties in the comparison.  For the \wmap\ chains, the value of $D_{220}$ was reported in the chains, so we adopt those values.  For the \planck\ chains we evaluate $D_{220}$ from the cosmological parameters using CAMB, including the Gaussian lensing approximation but excluding all foreground contributions (SZ effect, galactic foregrounds, etc.) since we seek to compare the CMB component of the power spectrum returned by each experiment.

From the first 10,000 points in each \planck\ chain we collect every 10th point to obtain 1000 relatively independent samples.  Given the values of $\Omega_b h^2$, $\Omega_c h^2$, $H_0$, $A_{s,0.05}$, $n_s$, $\tau$, $Y_{\rm He,used}$, and $\Omega_\nu h^2$, we recalculate the power spectrum assuming one massive neutrino and 2.046 massless neutrinos.  We use the March 2013 version of CAMB\footnote{\url{http://camb.info/}}, matching the version used to generate these chains.  The mean and standard deviation are computed using the weights reported in the \planck\ chains.  

The results for $D_{220}$ are given in Table~\ref{tab:cl_220}: we find the \planck\ CMB spectrum to be significantly lower than \wmap's.   While there is some variation between the three \planck\ chains explored, the range is small.  Based on only 1000 samples, we expect the sampling error to be about 3\% of the quoted uncertainty, so 1 $\mu{\rm K}^2$ differences in the mean are not significant.  Furthermore, CAMB does not claim an accuracy of more than $\sim$0.1\%, so even 6 $\mu{\rm K}^2$ variations should not be considered highly significant.  As a cross check of our calculations, we extract $D_{220}$ from the best-fit spectra released with the chains and find values consistent with those we derived: $5747.6~\mu{\rm K}^2$ for the \wmap\ chain run by \planck, $5586.9~\mu{\rm K}^2$ for the \planck+(\wmap\ low-$l$ pol.) chain, $5588.4~\mu{\rm K}^2$ for \planck\ alone, and $5597.5~\mu{\rm K}^2$ for the \planck+$\tau$ prior chain.

\capstartfalse
\begin{deluxetable}{lll}
\tablecaption{Comparison of $D_{220}$\tablenotemark{a} values\label{tab:cl_220}}
\tablewidth{0pt}
\tablehead{\colhead{Data} & \colhead{$D_{220}$ [$\mu{\rm K}^2$] } & \colhead{Source}}
\startdata
\wmap\    & $ 5746 \pm 35 $ & \wmap9 \\
\wmap\    & $ 5745 \pm 36 $ & \wmap9, re-run by \planck\ \\
          &               &   \\
\planck\  & $ 5588 \pm 35 $ & \planck\tablenotemark{b} + \wmap9 low-$l$ pol. \\
\planck\  & $ 5587 \pm 36 $ & \planck\ alone\tablenotemark{b} \\ 
\planck\  & $ 5598 \pm 35 $ & \planck\  + $\tau$ prior\tablenotemark{c}
\enddata
\tablenotetext{a}{$D_{220} \equiv l(l+1)C_{l}/(2\pi) |_{l=220}$.}
\tablenotetext{b}{$2 \le l \le 2500$}
\tablenotetext{c}{$50 \le l \le 2500$}
\end{deluxetable}
\capstarttrue

\capstartfalse
\begin{deluxetable}{lcc}
\tablecaption{\wmap\ beam characterization regimes\label{tab:regimes}\tablenotemark{a}}
\tablewidth{0pt}
\tablehead{\colhead{Band} & \colhead{Cutoff} & \colhead{Transition Radius} \\
 \colhead{} & \colhead{(dBi / dB)} & \colhead{(degrees)}}
\startdata
K  & 2.0 / $-$45.0 & 7.0 \\
Ka & 3.0 / $-$46.4 & 5.5 \\
Q  & 5.0 / $-$46.3 & 5.0 \\
V  & 6.0 / $-$48.8 & 4.0 \\
W  & 9.0 / $-$48.8 & 3.5
\enddata
\tablenotetext{a}{From \citet{bennett/etal:2013}.}
\end{deluxetable}
\capstarttrue

\begin{figure*}[ht]
\begin{center}
\includegraphics[height=5in]{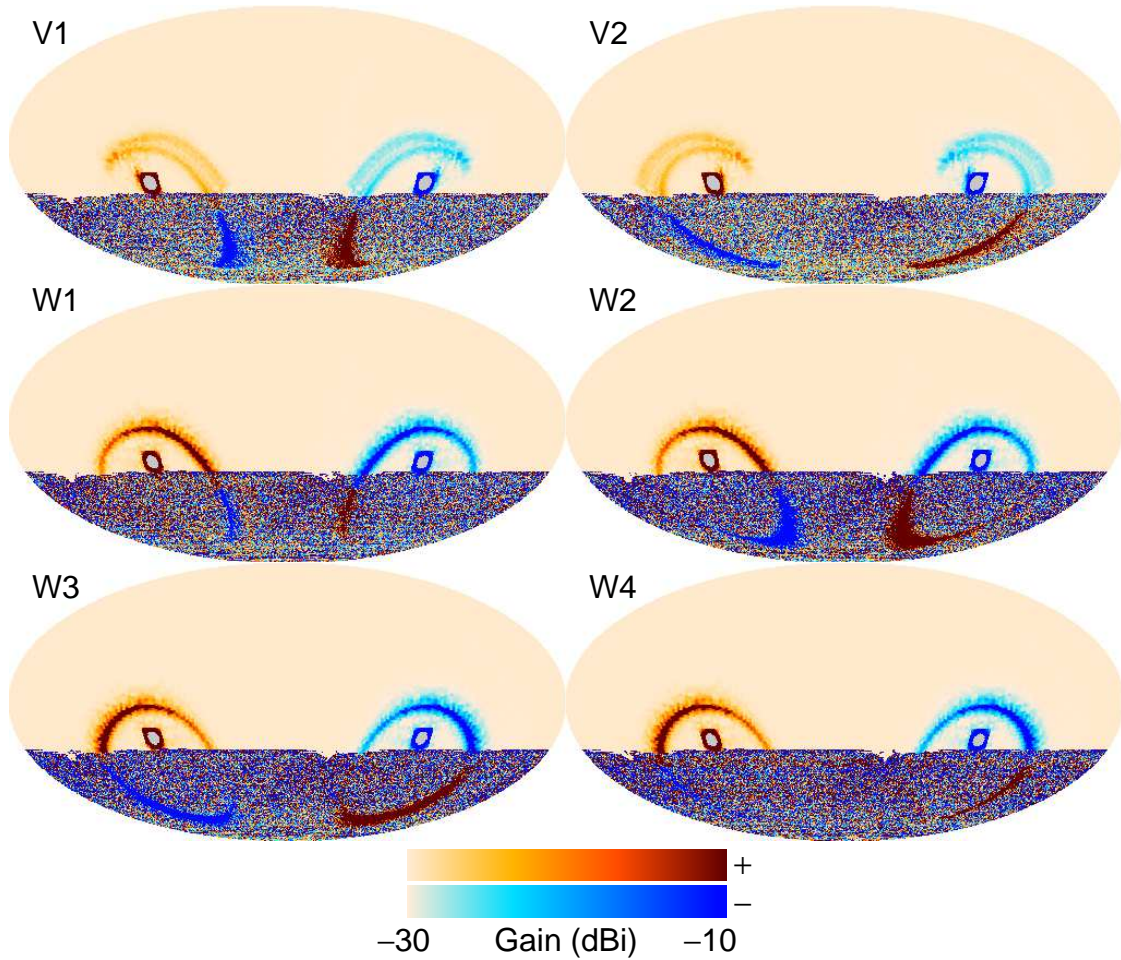}
\end{center}
\vspace{-0.1in}
\caption{\wmap\ far sidelobe beam maps for V1-W4 in spacecraft coordinates, where the top of the map is sunward and vice-versa.  The main beam and near sidelobe response, within the ``transition radius'', has been masked in grey and lies $\sim 20^\circ$ below the mid-plane of the map.  Positive response (from the A side) is shaded brown, and negative response (from the B side) is shaded blue.  The color scale is in linear gain units, $G=10^{({\rm dBi}/10)}$ and it saturates at $-10$ dBi, far below the forward gain of $55-58$ dBi, in V and W band. Creme indicates regions where $|G| < -30$ dBi.   The prominent rings on either side arise from radiation that spills past the edge of the primary reflector; the sign change in that response depends on whether or not such radiation also reflects off the large instrument radiator.  \label{fig:VW_sidelobe}}
\end{figure*}

If the \wmap\ and \planck\ errors were uncorrelated, the \wmap\ $D_{220}$ would be 3.2$\sigma$ higher than \planck's.  Since these chain-based uncertainties include calibration error (which will be correlated because \planck\ used the \wmap\ solar dipole calibration) and cosmic variance error (which will be correlated between \wmap\ and \planck\ due to overlapping sky coverage), we expect some correlation in the $D_{220}$ errors.  Specifically, the uncertainty in the {\em difference} between the \wmap\ and \planck\ will be lower than if the errors were uncorrelated; this will increase the significance of the discrepancy.  Using the simulations presented in \S\ref{sec:param_compare} we estimate that the $D_{220}$ uncertainties have a Pearson correlation coefficient of $r=0.69$, not including $l$-to-$l$ correlations in the window function uncertainties.  If we include this value of $r$, we estimate the difference, $\Delta D_{220}$, is  $5.7\sigma$ rather than $3.2\sigma$, but this probably over-estimates the significance slightly due to the window function treatment.  In the end, we estimate that the first peak amplitude is discrepant at somewhere between 3 and 5$\sigma$ significance.

Plausible explanations for the \planck\ $-$ \wmap\ spectrum difference include an overall brightness scale difference and/or beam characterization errors.  However the former would be puzzling because the \planck\ brightness calibration was based on the \wmap\ dipole and, as discussed in \S\ref{sec:map_compare}, the measured dipoles are relatively consistent.   \citet{hazra/shafieloo:2014b} find that a rescaled \planck\ concordance model is consistent with the \wmap\ data for $l < 1200$, but it does not follow that the measured discrepancy is actually due to a mis-calibration of either experiment's brightness scale.  We note that the recent \planck\ consistency paper \citep{planck/31:2013} claims some $l$ dependence in the \planck/\wmap\ ratio and suggests that the cause is not simply a calibration error.  Further, that paper reports a partial reduction in the \planck/\wmap\ spectrum difference following a revised analysis of the \planck\ beam response.   Since the Collaboration has not yet released a new likelihood code or Markov chains, we cannot determine the effect it has on $D_{220}$.  Further, given the restricted scope of the first \Planck\ data release, we cannot independently assess the \Planck\ Collaboration's estimates of systematic errors.  Instead, we revisit potential sources of systematic error in the \wmap\ beam analysis, especially the possibility that the \wmap\ solid angle has been systematically over-estimated, as would be required to explain the sign of the spectrum discrepancy.

\subsection{\wmap\ beam uncertainties revisited}\label{sec:wmap_beam}

For the purposes of \wmap\ data processing, the treatment of beam effects is divided into three regimes \citep{hill/etal:2009}, each of which are characterized and processed differently:\newline

\emph{Main beam}:
In this regime, the beam response is determined entirely by in-flight measurements of Jupiter.  The extent of the main beam is defined by an amplitude cutoff, measured in dBi.  See Table~\ref{tab:regimes} for the band-specific cutoff values.  \newline

\emph{Near sidelobe}:
In this regime, the response is determined by a physical optics model fit to in-flight measurements of Jupiter.  The fit is performed simultaneously to all data channels.  The outer limits of this regime are defined by a fixed ``transition radius'' for each band (see Table~\ref{tab:regimes}), while the inner limit is defined by the amplitude cutoff noted above.  In this regime the beam is characterized by a mixture of Jupiter data and the physical model, governed by the amplitude of the model. \newline

\emph{Far sidelobe}:
In the first two regimes, the A-side and B-side beams were determined separately and treated as independent antennas.  In the far sidelobe regime (the full sphere outside the transition radius of each beam) the response is treated as a proper differential antenna which carries a sign.  The data informing the response comes from four sources, listed in order of importance: 1) ground-based measurements of the beam pedestal (within ~$10^\circ$) using the reflector evaluation unit (REU) in the Goddard Electromagnetic Anechoic Chamber (GEMAC) at the Goddard Space Flight Center, 2) in-flight measurements of the response using the moon as a source, 3) measurements taken on the roof of Jadwin Hall at Princeton University using the REU, and 4) a physical optics model computed with the code DADRA.  The compiled far sidelobe maps are shown in Figure~\ref{fig:VW_sidelobe}, based on Fig.~2 of \cite{barnes/etal:2003}.  We estimate the uncertainty of the higher gain regions to be $\sim$30\%, based on in-flight measurements of the moon using in-flight calibration data, and matching this to the GEMAC data. \newline

In the data processing, the regime inside the transition radius (main beam and near sidelobe) is treated as a point spread function (PSF) with a radial profile which treats the beam as circularly symmetric.  This PSF convolves the angular power spectrum with a beam window function given by the Legendre transform of the radial profile.  Appendix B of \cite{hinshaw/etal:2007} analyzed the effects of beam asymmetry on the convolution and concluded that for \wmap's scan strategy and sensitivity levels, the effects of beam asymmetry on power spectrum convolution were $<$1\%, in power units, for V and W bands.  This topic was revisited in \cite{bennett/etal:2013} with the development of an explicit map-making method that deconvolved the effects of beam asymmetry, see especially Figure~9 and related text.  This work upheld the previous conclusions quoted above.

The \wmap\ processing pipeline treats the far sidelobe regime differently in that it estimates the differential far sidelobe contribution for each time-ordered data point using a model sky map based on the data.  This correction is incorporated into the intensity calibration algorithm so that it self-consistently accounts for the $l=1$ (dipole) response of the far sidelobe when estimating the intensity scale.  We describe this processing step more fully later in this section.

\begin{figure}[ht]
\begin{center}
\vspace{-0.15in}
\includegraphics[width=0.50\textwidth]{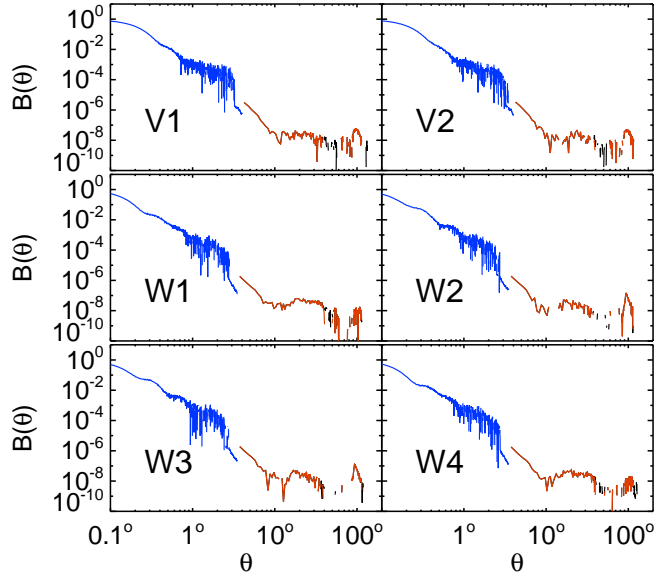}
\vspace{-0.5in}
\end{center}
\caption{The azimuthally-averaged radial profile of the \wmap\ beam response for the A-side V and W band differencing assemblies, normalized to a peak value of 1.  The blue portion indicates the main beam and near sidelobe response within the transition radius; the black and red portions are two representations of the (differential) far sidelobe response binned by distance from the A-side boresight.  The red trace includes all structure shown in Figure~\ref{fig:VW_sidelobe}, irrespective of sign, while the black trace excludes spill past the B-side reflector, as described in the text and as shown in Figure~\ref{fig:beam_digest}.  In the nominal \wmap\ data processing, the sidelobe response is corrected for in the differential data, prior to map-making. \label{fig:profile_A}}
\vspace{0.1in}
\end{figure}

\begin{figure}[ht]
\begin{center}
\includegraphics[width=0.5\textwidth]{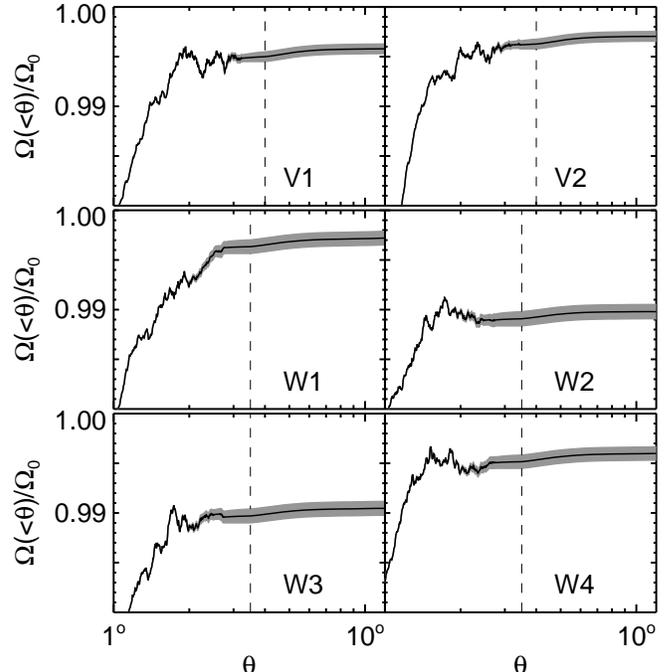}
\vspace{-0.5in}
\end{center}
\caption{Cumulative solid angle of the A-side V and W band beam profiles, normalized to 1 at 180$^{\circ}$.  The dashed line indicates the transition radius, inside of which the \wmap\ beams are modeled using a combination of in-flight Jupiter data and a physical optics model.  With the exception of W2 and W3, the response within the transition radius accounts for $\sim$99.5\% of the beam solid angle.  The grey band shows the effects of altering the amplitude of the model contribution by $\pm$100\%: the cumulative solid angle changes by $\sim$0.1\%.  Results for the B side are similar.  The normalization at 180$^{\circ}$ is established using the masked sidelobe response (black trace in Figure~\ref{fig:profile_A}); we explore the unmasked treatment in Figure~\ref{fig:beam_digest}.  In the nominal \wmap\ data processing, the far sidelobe response (outside the transition radius) is corrected for in the differential data, prior to map-making.  \label{fig:csa_012_and_sl}}
\vspace{0.1in}
\end{figure}

\subsubsection{Beam profiles and solid angle estimates}\label{sec:re_beam}

As noted above, the far sidelobe response is treated as a differential signal correction in the \wmap\ data processing.  However, for study purposes it is useful to estimate the fraction of the  beam solid angle that is attributable to the far sidelobes.  This attribution is not unambiguous because certain far sidelobe features cross the mid-plane dividing the A and B sides of \wmap\, and thus complicate a naive accounting of the solid angle.  Nonetheless, we can obtain useful insights and so we proceed.

Figure~\ref{fig:profile_A} shows the azimuthally averaged radial profile of the 6 V and W-band A-side DA's, compiled from the sources listed above.  The blue portion indicates the main beam and near sidelobe: the regime that is accounted for using a window function.   The red portion shows the result of averaging the far sidelobe response (Figure~\ref{fig:VW_sidelobe}) in radial bins centered on the A-side boresight.  The black curve is a modified form of the red curve where unambiguous negative B-side contributions have been masked from the far sidelobe response to assess the magnitude of their contribution.  Figure~\ref{fig:csa_012_and_sl} shows the fractional cumulative solid angle derived from these profiles as a function of distance from the A-side boresight.  These integrals are evaluated all the way out to $180^\circ$ for study purposes, but they are only used inside the transition radius ($3.5^\circ$ and $4.0^\circ$ for W and V bands, respectively).  We note the following items in Figure~\ref{fig:csa_012_and_sl}:
\begin{enumerate}
\item The fraction of solid angle contributed by the far sidelobe response (outside the transition radius) is between 0.5\% and 1\%, but this result does not fully account for the differential nature of the far sidelobe response.  Beyond about $\sim$20$^{\circ}$ from each boresight, we begin to mix A-side (positive) and B-side (negative) response due to spill past the primary.  We explore alternative accountings below.
In the nominal \wmap\ data processing, the sidelobe response is corrected for in the differential data, prior to map-making.  We revisit that procedure below as well.
\item The fraction of solid angle in the response between the transition radius and $20^\circ$ is at most 0.1\% for any DA.  This limits the plausible range over which errors in this response could be affecting the far sidelobe correction.
\item The solid angle in the region between $\sim1^{\circ}$ and the transition radius arises from a combination of Jupiter data and a physical optics model.  The uncertainty in the model contribution is indicated by the grey band which shows the affect of varying the amplitude of the model by $\pm$100\%: the solid angle changes by $\sim$0.1\%.  In particular, setting the model contribution to zero only reduces the  beam solid angle within the transition radius by 0.1\%.
\item The remainder of the beam solid angle within the transition radius is characterized by in-flight measurements of Jupiter.  These data and their uncertainties are very well understood and are reflected in the published \wmap\ beam covariance matrices.
\end{enumerate}

Recall that \wmap's measured power spectrum is higher than that reported by \planck.  One way to reduce the measured spectrum is to reduce the estimated beam solid angle, which reduces the factor by which one deconvolves the raw spectrum.  Since the physical optics model is contributing at most 0.1\% to the overall solid angle, this limits the degree to which the final power spectrum amplitude can be reduced ($\sim$0.2\%).  We must look to the far sidelobes, especially item 1 above, for any remaining beam-related sources of error, beyond what is already published in the \wmap\ beam covariance matrices.

\subsubsection{\wmap\ far sidelobe response and calibration}\label{sec:re_sidelobe}

As discussed earlier in \S\ref{sec:wmap_beam}, the far sidelobe response maps are compiled from a variety of sources \citep{barnes/etal:2003, hill/etal:2009}, see Figure~\ref{fig:VW_sidelobe}.  They are dominated by power at $l < 40$ with the main features being a pedestal from the transition radius out to $\sim 20^{\circ}$, and direct spill past the primary mirror.  We estimate an overall uncertainty in the highest-gain features in these maps to be about 30\%.  

An estimate of the far sidelobe response is removed in the time-ordered data as follows \citep{hinshaw/etal:2009}: we evaluate a hyper-cube of responses to a dipole and a model sky map (based on earlier data) for all possible spacecraft Euler angles, with $\sim$1$^{\circ}$ spacing.  For each time-ordered data point, we evaluate and apply the correction, in raw counts, prior to evaluating a dipole-based gain solution, so that our response to the dipole is, in principle, only convolved by the beam response within the transition radius.  We have evaluated the overall gain solutions with and without applying this correction and have tabulated the differences between them in Table~\ref{tab:recal}.  For five-year \wmap\ data release and beyond, the gain corrections range between $-0.05$\% and $+0.23$\% for V and W band.  For reference, Table~\ref{tab:recal} also gives the factors for the three-year release, which used a slightly different transition radius and treatment of the sidelobe correction.  The results are very similar, but generally a little larger. 

In the vicinity of first acoustic peak, where the power spectrum discrepancy is most significant, the \wmap\ V and W band data contribute roughly equally to the combined CMB spectrum.  In that regime, we estimate the net recalibration applied to the combined data is $\sim$0.08\%.   The far sidelobe response uncertainty of 30\% implies a comparable fractional uncertainty in the net recalibration factor.  Even if we were to inflate the uncertainty to 100\%, corrections to the recalibration would be insufficient to close the observed gap between the \wmap\ and \planck\ spectra.

\begin{figure}[ht]
\includegraphics[width=0.52\textwidth]{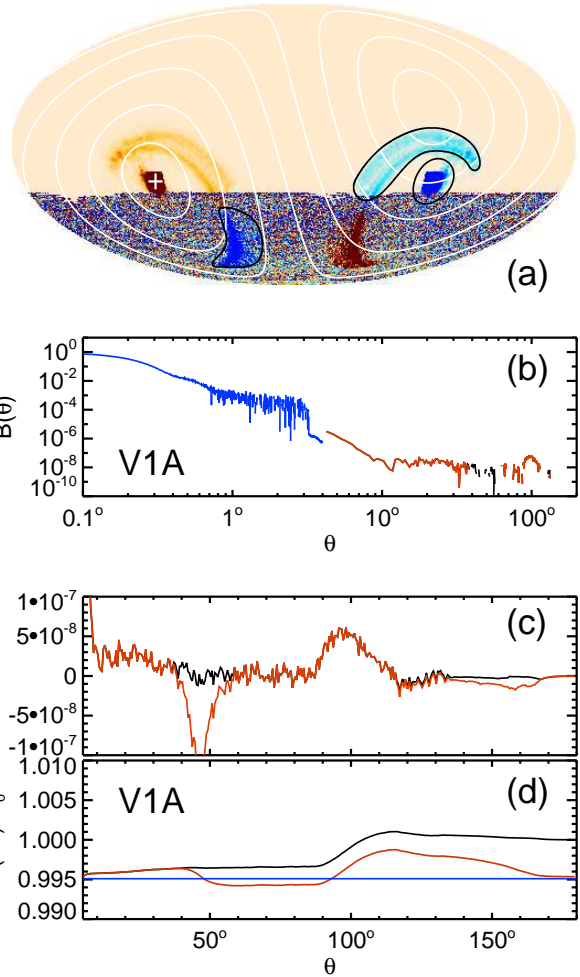}
\vspace{-0.3in}
\caption{A summary of far sidelobe tests applied to the \wmap\ V1 differencing assembly.  (a) The full-sky differential beam map, as in Figure~\ref{fig:VW_sidelobe} but with the main beam and near sidelobe regions unmasked.  The white cross indicates the centroid of the A-side beam, and lines of constant angular distance from the A-side centroid are shown as white circles, in $20^\circ$ increments.  Regions outlined in black indicate relatively bright sidelobes of the B-side beam that are optionally masked when analyzing the A-side solid angle.  (b) The azimuthally-averaged beam response, as in Figure~\ref{fig:profile_A}.  (c) Zoom of (b) in linear units. This illustrates the effect of including (red) or excluding (black) the B-side primary spill (outlined region) in the A-side solid angle computation.  The response within $12^\circ$ of the B-side beam centroid is excluded in both cases.  (d) The cumulative solid angle at large angles, normalized to 1 at $\theta = 180^{\circ}$.  The nominal treatment excludes the B-side primary spill (black trace); the red trace indicates that including the negative B-side spill would reduce the A-side solid angle by 0.5\%.  99.5\% of the solid angle is contributed by the main beam and near sidelobe, as shown in blue.  Very large fractional and conspiring errors would be required to produce a significant change in our estimate of the solid angle from the far sidelobe.  See the text for further discussion. \label{fig:beam_digest}}
\end{figure}

\capstartfalse
\begin{deluxetable}{lcc}
\tablecaption{\wmap\ far sidelobe recalibration factors \label{tab:recal}}
\tablewidth{0pt}
\tablehead{\colhead{Band/} & \colhead{\wmap5+\tablenotemark{a}} & \colhead{\wmap3\tablenotemark{b}} \\
 \colhead{sub-band} & \colhead{(\%)} & \colhead{(\%)}}
\startdata
V    & $+$0.05 & $+$0.10 \\
W14  & $+$0.23 & $+$0.44 \\
W23  & $-$0.05 & $-$0.12
\enddata
\tablenotetext{a}{\wmap\ five-, seven- and nine-year data releases \citep{hinshaw/etal:2009}.}
\tablenotetext{b}{\wmap\ three-year release \citep{jarosik/etal:2007}.}
\vspace{0.1in}
\end{deluxetable}
\capstarttrue

\section{Cosmological parameter comparison}\label{sec:param_compare}

In this section we consider the standard 6-parameter \lcdm\ model and compare the best-fit parameters obtained from the \wmap\ nine-year data (alone) and the initial \planck\ data plus the \wmap\ polarization data (``\WMAP{pol}'').  The best-fit parameters for each data set are given in Table~\ref{tab:param_comp}, while the best-fit power spectra derived from each model are shown and compared in Figure~\ref{fig:planck_wmap_powspec_ratio}.  

\capstartfalse
\begin{deluxetable*}{lccccr}
\tablecaption{Mean and maximum likelihood cosmological parameters\label{tab:param_comp}}
\tablewidth{0pt}
\tablehead{\colhead{Parameter} & 
\colhead{\wmap\ mean} & 
\colhead{\Planck\ mean\tablenotemark{a}} &
\colhead{\wmap\ m.l.\tablenotemark{b}} & 
\colhead{\Planck\ m.l.\tablenotemark{c}} &
\colhead{\wmap$-$\planck\ m.l.\tablenotemark{d}}}
\startdata
$\Omega_b h^2$                     & $0.02264 \pm 0.00050$ & $0.02205 \pm 0.00028$     & 0.02262 & 0.02194 & $0.00068\pm 0.00050\ \  (1.4\sigma)$ \\
$\Omega_c h^2$                     & $0.1138 \pm 0.0045$   & $0.1199 \pm 0.0027$       & 0.1139  & 0.1220  & $-0.0081\pm 0.0049\ \ \ \,(1.7\sigma)$\\
$H_0$ [km s$^{-1}$ Mpc$^{-1}]$     & $70.0 \pm 2.2$        & $67.3 \pm 1.2$            & 70.0    & 66.8    & $3.2\pm 2.3\qquad\ \ (1.4\sigma)$ \\
$n_s$                              & $0.972\pm 0.013$      & $0.9603\pm 0.0073$        & 0.974   & 0.953   & $0.020\pm 0.014\quad\ \ (1.4\sigma)$ \\
$10^9 A_{s,0.05}$\tablenotemark{e} & $2.203 \pm 0.067$     & $2.196\pm 0.055$          & 2.203   & 2.200   & $0.004\pm 0.042\quad\ \ (0.1\sigma)$ \\
$\tau$                             & $0.089 \pm 0.014$     & $0.089^{+0.012}_{-0.014}$ & 0.0875  & 0.0873  & $0.0002\pm 0.0031\ \ \ \,   (0.1\sigma)$ 
\enddata
\tablenotetext{a}{``\Planck+WP'' results, from Table~2 of \citet{planck/16:2013}.  This uses \planck\ data for $2\le l\le 2500$,
as well as the low-$l$ \wmap\ polarization likelihood (WP).}
\tablenotetext{b}{The \wmap\ maximum likelihood parameters were determined by fitting a quadratic function to the peak of the log likelihood, based on the released \wmap9 \lcdm\ chain. The SZ amplitude was constrained to be positive.}
\tablenotetext{c}{The \Planck\ maximum likelihood parameters were computed by maximizing the \Planck\ TT likelihood code ($50 \le l \le 2500$) over 20 parameters: the 6 \lcdm\ parameters and 14 foreground and calibration parameters.  We assumed an optical depth prior of $\tau=0.085\pm0.01$.}
\tablenotetext{d}{Single-parameter differences with marginalized uncertainties.  The maximum likelihood parameters are used and the quoted errors account for correlations between the experiments using the rescaled simulation covariance matrix discussed in the text.}
\tablenotetext{e}{The \wmap\ team reports the spectrum normalization at $k_0=0.002$ Mpc$^{-1}$ while the \planck\ Collaboration adopts $k_0=0.05$ Mpc$^{-1}$.  We report $A_{s,0.05}$, calculated from released chains, and we note that $A_{s,0.002} = A_{s,0.05}\times 25^{1-n_s}$ 
for the simple \lcdm\ model without a running spectral index.}
\tablecomments{The results in the column ``\planck\ mean'' assume a non-zero neutrino mass, as per  \citet{planck/16:2013}; the remaining columns set the neutrino mass sum to 0 and the helium fraction to $Y_{\rm He}=0.24$.}
\end{deluxetable*}
\capstarttrue

At the level of single parameters, the inferred values are generally consistent within their (marginalized) errors.  For example, \wmap\ estimates the Hubble constant to be $H_0 = 70.0 \pm 2.2$ km s$^{-1}$ Mpc$^{-1}$ while \planck\ estimates $H_0 = 67.3 \pm 1.2$ km s$^{-1}$ Mpc$^{-1}$ \citep{bennett/etal:2014} If these errors were uncorrelated and Gaussian, we would expect the measurements to differ by 2.5 km s$^{-1}$ Mpc$^{-1}$ (1$\sigma$), making the observed difference 1.1 $\sigma$.  However, this comparison ignores the fact that a substantial fraction of the uncertainty in each parameter is due to cosmic variance, which is common between the two experiments; thus it over-estimates the consistency.  (Looking ahead, the final column of Table~\ref{tab:param_comp} quotes single parameter differences with uncertainties that account for these correlations, using the methodology described below.)  More importantly, a complete test of parameter consistency requires that we simultaneously assess all 6 \lcdm\ parameter differences accounting for their parameter covariance.

A simple test of 6-dimensional parameter consistency is to examine how much ``overlap'' exists between points in a \wmap\ parameter chain and a corresponding \planck\ chain.   It turns out that a single hyper-plane in parameter space segregates the chain points from each experiment at the 0.1\% level.  Specifically, the parameter combination
\begin{multline}
P \equiv \Omega_b h^2 -0.874789 \,\Omega_c h^2 -0.000736964 \,H_0 \\
-0.051314 \,n_s +0.0710558 \times 10^9 A_{s,0.05} \\
-0.316342 \,\tau + 0.052037
\label{eq:maxdiff}
\end{multline}
where $H_0$ is in km s$^{-1}$ Mpc$^{-1}$ and the other parameters are dimensionless, splits the experiments as follows: $<$~0.1\% of the \planck-released \wmap\ chain weight has $P<0$, while $<$~0.1\% of the \planck+\wmap{pol} chain weight has $P>0$.

\subsection{Simulating experiment covariance}\label{sec:exp_cov}

We next proceed to quantify the covariance expected between the parameters inferred from each experiment.  To do so, we generate simulations that approximate the full parameter recovery process for each experiment and we use them to generate the full covariance matrix of parameter differences expected given the effects of independent instrument noise, different sky masks, and different multipole ranges probed by the two experiments.

To make the problem computationally tractable, we do not perform a full Markov chain analysis for each simulation realization but rather we estimate the most likely parameter value for each realization (and data set), using the {\tt scipy} implementation of the Powell method.  We then compare the measured parameter differences to the statistical distribution of simulated parameter differences to assess the consistency of the two measured parameter sets.  To further speed up the calculation we ignore 14 calibration and foreground nuisance parameters used in the full \Planck\ likelihood.  Since our simulated maps do not include any modeled foreground residuals or calibration artifacts, ignoring these parameters will not introduce any bias.

The simulation program proceeds as follows: for each realization in our simulation, we generate a CMB map from a fixed \lcdm\ power spectrum whose parameters are given in Table~\ref{tab:param_comp}.  We compute the $a_{lm}$ coefficients of this (masked) map using PolSpice\footnote{\url{http://www2.iap.fr/users/hivon/software/PolSpice/}} \citep{szapudi/etal:2001,chon/etal:2004}; for the \wmap\ simulation we apply the KQ85y9  temperature analysis mask \citep{bennett/etal:2013} while for \planck\ we use the CL49 temperature mask for $50 \le l \le 1200$ and the CL31 mask for $l > 1200$ \citep{planck/15:2013}.   We add white noise to each spectrum as described below, then we form a simple Gaussian, diagonal likelihood for each experiment using,
\beq
-2 \ln \mathcal{L} = \chi^2 = \sum_{l=l_{\rm min}}^{l_{\rm max}} \frac{(C_l^{\rm sim} - C_l^{\rm th})^2}{\sigma_l^2},
\label{eq:diag_like}
\eeq
where $C_l^{\rm sim}$ is the simulated $C_l$ realization, $C_l^{\rm th}$ is the model spectrum to be fit, and $\sigma_l^2$ is the variance per $l$.  For \wmap\ we take $32\le l \le 1200$ and for \planck\ we take $50 \le l \le 2500$.  In addition to equation~\ref{eq:diag_like}, we account for low-$l$ polarization data by imposing a Gaussian prior on $\tau$, as discussed in more detail below.  In the end, we generate a database of $2 \times 2000$ maximum-likelihood parameters for each experiment and simulation realization. 

Before discussing the results, we describe the construction of our (diagonal) power spectrum covariance matrix in more detail, and establish that it provides an adequate approximation to the full likelihood function for each experiment.  The form of our matrix neglects mask-induced anti-correlations between nearby $l$ values, but since the theory power spectra, $C_l^{\rm th}$, are smooth functions of $l$, the effect of the off-diagonal elements can be approximated by summing the variance contributions in the neighborhood of each diagonal element.  Given the full covariance matrix, $N_{ll'}$ for a given experiment, we can form the approximate diagonal matrix as
\be
\sigma_l^2 = \sum_{i} N_{l-i,l+i}
+ \frac12 \sum_{i} N_{l-i+1,l+i} 
+ \frac12 \sum_{i} N_{l-i,l+i+1}.
\label{eq:smooth_N}
\ee
For \wmap, the full covariance is taken from the nine-year \wmap\ likelihood code and the sum over $i$ runs over the full extent of the matrix (excluding $l=0,1$). For \planck, we form a frequency-averaged, foreground-removed covariance matrix from the full matrix provided in their likelihood code (described in Appendix~\ref{sec:simplify_planck}), and we restrict the sum to $-10 \le i \le +10$ for the first term, $-9 \le i \le +10$ for the second term, and $-10 \le i \le +9$ for the last term.

This approximate diagonal matrix correctly accounts for off-diagonal contributions to the variance in the limit that structure in the spectra are smooth.  In order to determine the self-consistent level of instrument noise to add to each simulated spectrum, we estimate the cosmic variance contribution, $\sigma_{l,{\rm cv}}^2$, then add Gaussian random noise with variance $\sigma_{l,{\rm n}}^2 = \sigma_l^2 - \sigma_{l,{\rm cv}}^2 \ge 0$.  The effective cosmic variance is determined by filtering each simulated spectrum with a $\Delta l=21$ boxcar filter (the smoothing scale applied in equation~\ref{eq:smooth_N}), multiplying by $\sqrt{21}$ to correct for the $rms$ suppression induced by smoothing, then taking a standard deviation at each $l$ over all 2000 realizations for each experiment. 

Because the \planck\ Collaboration has not yet completed their low-$l$ polarization analysis, they utilize \wmap\ polarization maps to constrain the optical depth, $\tau$.  As a result, the Gaussian prior we impose has the same peak for both experiments, so our $\tau$ statistics are highly correlated between the two.  Further, since we use only temperature data in our approximate likelihood, the simulations only constrain the combination $A_s \exp(-2\tau)$, and not $A_s$ or $\tau$ independently.  As a result, we find that the recovered value of $\tau$ in any given realization is essentially set by the peak of the input $\tau$ prior.  In order to mimic the true uncertainty in $\tau$, while avoiding the (costly) use of low-$l$ polarization data, we vary the $\tau$ prior applied to both likelihoods in each realization by randomly selecting the peak value from a Gaussian distribution with a mean value of 0.0851 and a standard deviation of 0.013.  Note that the width of our prior is unimportant because we always recover the peak value, but the scatter from one realization to the next correctly approximates the true uncertainty for each experiment.

The diagonal likelihood function is rapidly evaluated once $C_l^{\rm sim}$ and $C_l^{\rm th}$ are known.  However, we require a concomitant speedup in the evaluation of $C_l^{\rm th}({\bf p})$ given the cosmological parameters ${\bf p}$ in order to make the Monte Carlo-based comparison feasible.  To achieve this, we use a custom python implementation of the PICO algorithm which casts $C_l^{\rm th}({\bf p})$ as a a 4th-order polynomial within a specified volume of ${\bf p}$ around the best fit region.  We describe the details of the implementation in Appendix~\ref{app:pico} and note here that our polynomial form of $C_l^{\rm th}({\bf p})$ was found to be within 0.09 $\mu$K$^2$ (median absolute deviation) of the spectrum evaluated with CAMB.

\begin{figure*}[htp]
\begin{center}
\includegraphics[height=4in]{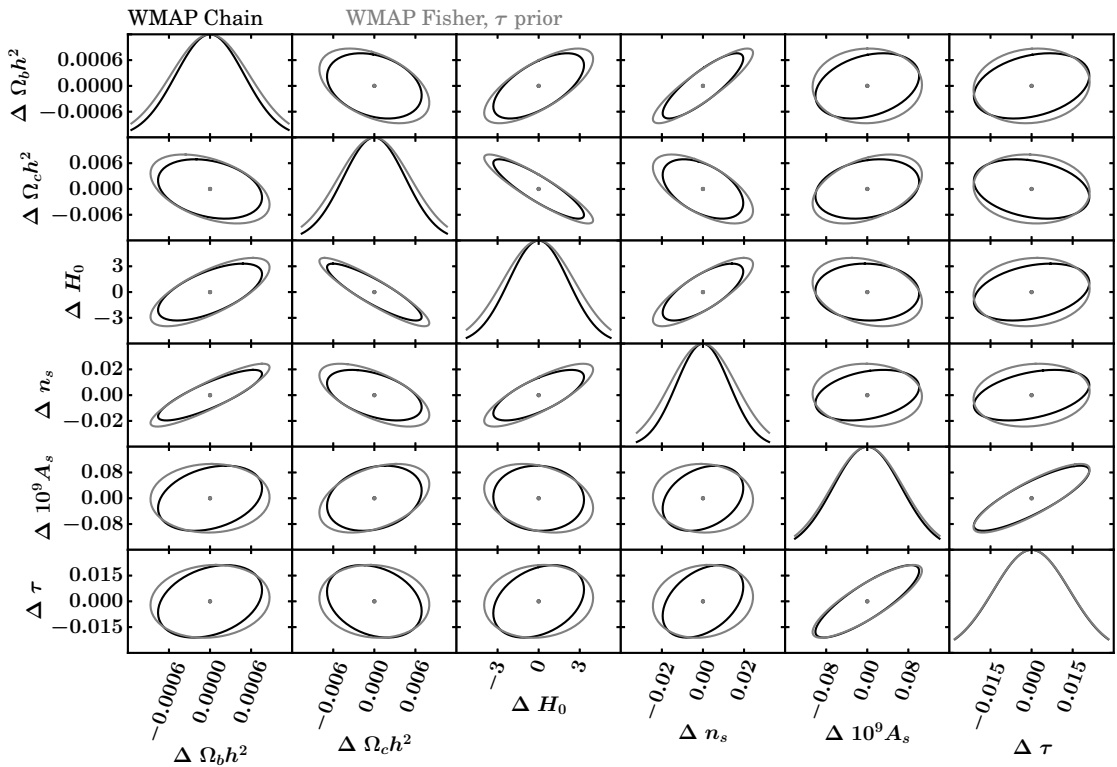}
\end{center}
\vspace{-0.15in}
\caption{\wmap\ parameter uncertainties derived from the full \wmap\ \lcdm\ chain (black) and from a Fisher analysis of the approximate diagonal \wmap\ likelihood used in this paper (grey).  The approximate likelihood includes a $\tau$ prior with scatter $\sigma_\tau$ chosen to mimic the \wmap\ low $l$ polarization likelihood.  The scalar amplitude $A_s$ is measured at $k_0 = 0.05$ Mpc$^{-1}$ and the Hubble constant $H_0$ has units of km s$^{-1}$ Mpc$^{-1}$.  Note that the projected simulation uncertainties are slightly larger than those derived from the full likelihood, hence these uncertainties are likely conservative.
\label{fig:wmap_sim}}
\end{figure*}

\begin{figure*}[ht]
\begin{center}
\includegraphics[height=4in]{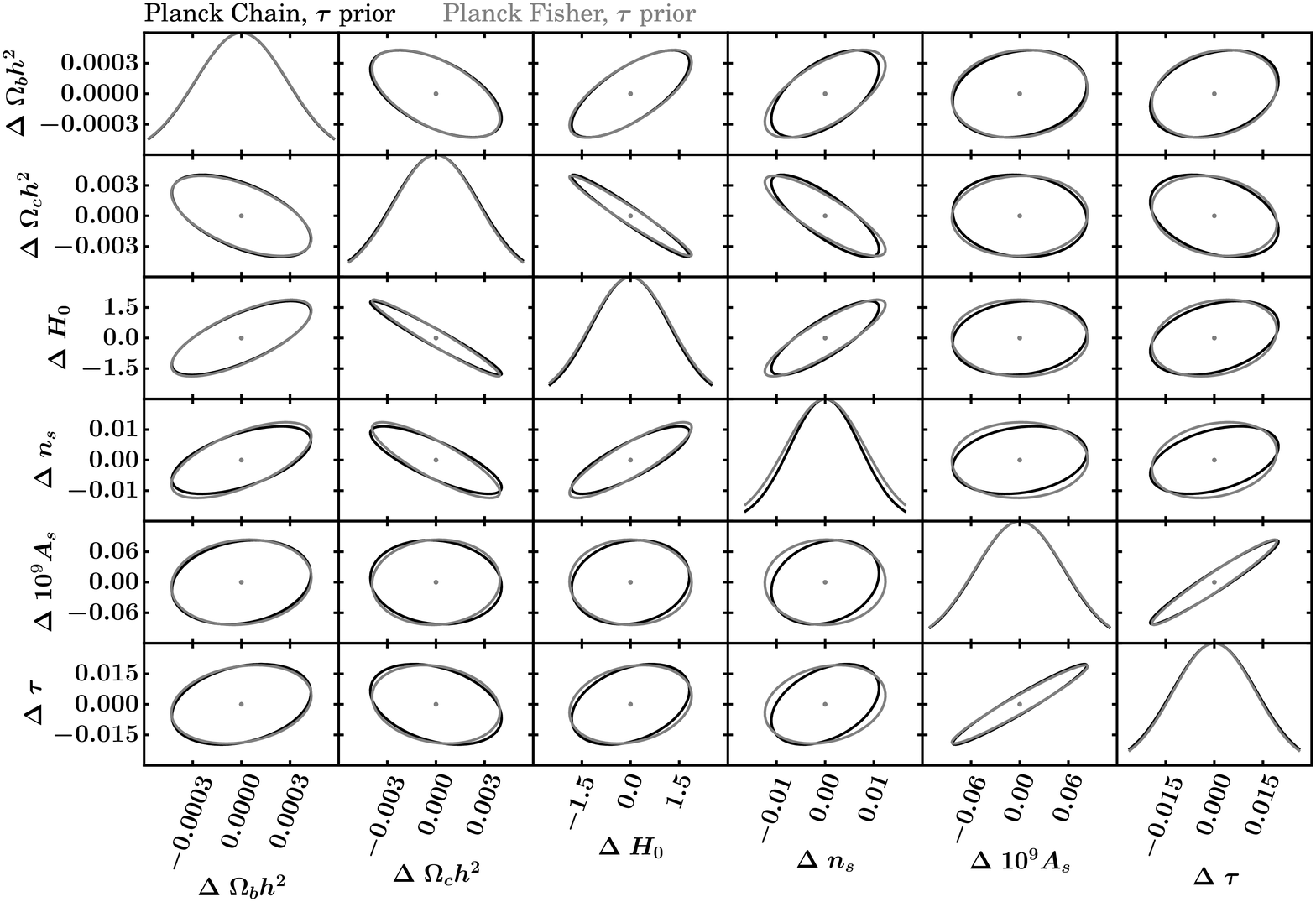}
\end{center}
\vspace{-0.15in}
\caption{\planck\ parameter uncertainties derived from the full \planck+\wmap{pol} \lcdm\ chain ``{\tt PLA/base/planck\_lowl\_lowLike}'' (black) and from a Fisher analysis of the approximate diagonal \planck\ likelihood used in this paper (grey).  The Fisher analysis includes a $\tau$ prior with scatter $\sigma_\tau =0.013$ to mimic the \wmap\ low $l$ polarization likelihood used in the chain, and it also includes marginalization over the thermal SZ template provided in the \Planck\ likelihood. The scalar amplitude $A_s$ is measured at $k_0 = 0.05$ Mpc$^{-1}$ and the Hubble constant $H_0$ has units of km s$^{-1}$ Mpc$^{-1}$.  The approximate uncertainties we derive from the diagonal likelihood agree closely with those obtained from the full \planck\ chain, showing that the diagonal likelihood is a useful approximation to the full likelihood.
\label{fig:planck_sim}}
\end{figure*}

\subsection{Simulation testing}\label{sec:sim_test}

Figures~\ref{fig:wmap_sim} and \ref{fig:planck_sim} provide a visualization of our simulation fidelity by comparing our simulated parameter covariance matrices to those derived from the flight data, for \wmap\ and \planck, respectively.  The black traces are derived from the Markov chains supplied by each experiment, the grey traces are derived from our simulations, as described below.  The diagonal panels show Gaussian PDFs with a variance given by the corresponding elements of the covariance matrix.  The off-diagonal panels show the 68\% confidence region of  the Gaussian PDF, marginalized over the four remaining parameters;  these panels highlight the correlations among the parameters.

The simulation covariances shown in Figures~\ref{fig:wmap_sim} and \ref{fig:planck_sim} were derived from a Fisher matrix analysis of the diagonal likelihood in Equation~\ref{eq:diag_like}.  Enumerating the cosmological parameters as $p_{\alpha}$ for $\alpha=1\ldots 6$, we form the matrix of spectrum derivatives,
\be
M_{\alpha l} \equiv \left. \frac{\partial C_l}{\partial p_\alpha}\right|_{p_\beta\; {\rm fixed}},
\ee
over the range of $l$ used in each likelihood, and then form the $6 \times 6$ inverse parameter covariance matrix as
\be
N^{-1}_{\alpha\beta} = M_{\alpha l} N^{-1}_{ll'} (M^T)_{l' \beta}.
\ee
The confidence regions derived from $N_{\alpha\beta}$ are shown in Figures~\ref{fig:wmap_sim} and \ref{fig:planck_sim} for \wmap\ and \planck.

To test the validity of our simulations further, we rescale the simulation covariance matrices as described below and evaluate the sensitivity of the inferred $\chi^2$ to our choice of rescaling.   We start by evaluating the full $12 \times 12$ covariance matrix, ${\bf C}$, of the combined \wmap\ and \Planck\ parameters, including the off-diagonal blocks that measure the correlations between the two.  We generate scaling factors by comparing the simulation variances to the data, $s_i^2 = C'_{ii}/C_{ii}$, as described below, and rescale the full covariance matrix according to $C'_{ij} = C_{ij}s_i s_j$.  To project this down to the parameter differences we form the $6 \times 6$ matrix, ${\bf \Delta C}'$ as
\be
{\bf \Delta C}' = \left( \begin{array}{cc} {\bf I} & -{\bf I} \end{array} \right)
\left( \begin{array}{cc} {\bf C}'_{WW} & {\bf C}'_{WP} \\ {\bf C}'_{PW} & {\bf C}'_{PP} \end{array} \right)
\left( \begin{array}{r} {\bf I} \\ -{\bf I} \end{array} \right),
\ee
where each term is a $6 \times 6$ matrix, ${\bf I}$ is the identity matrix, and the subscripts W and P refer to \wmap\ and \planck, respectively.  Results with and without this rescaling of the parameter difference covariance matrix are summarized in \S\ref{sec:param_result} below.

\subsection{Results}\label{sec:param_result}

\begin{figure*}
\begin{center}
\includegraphics[height=4in]{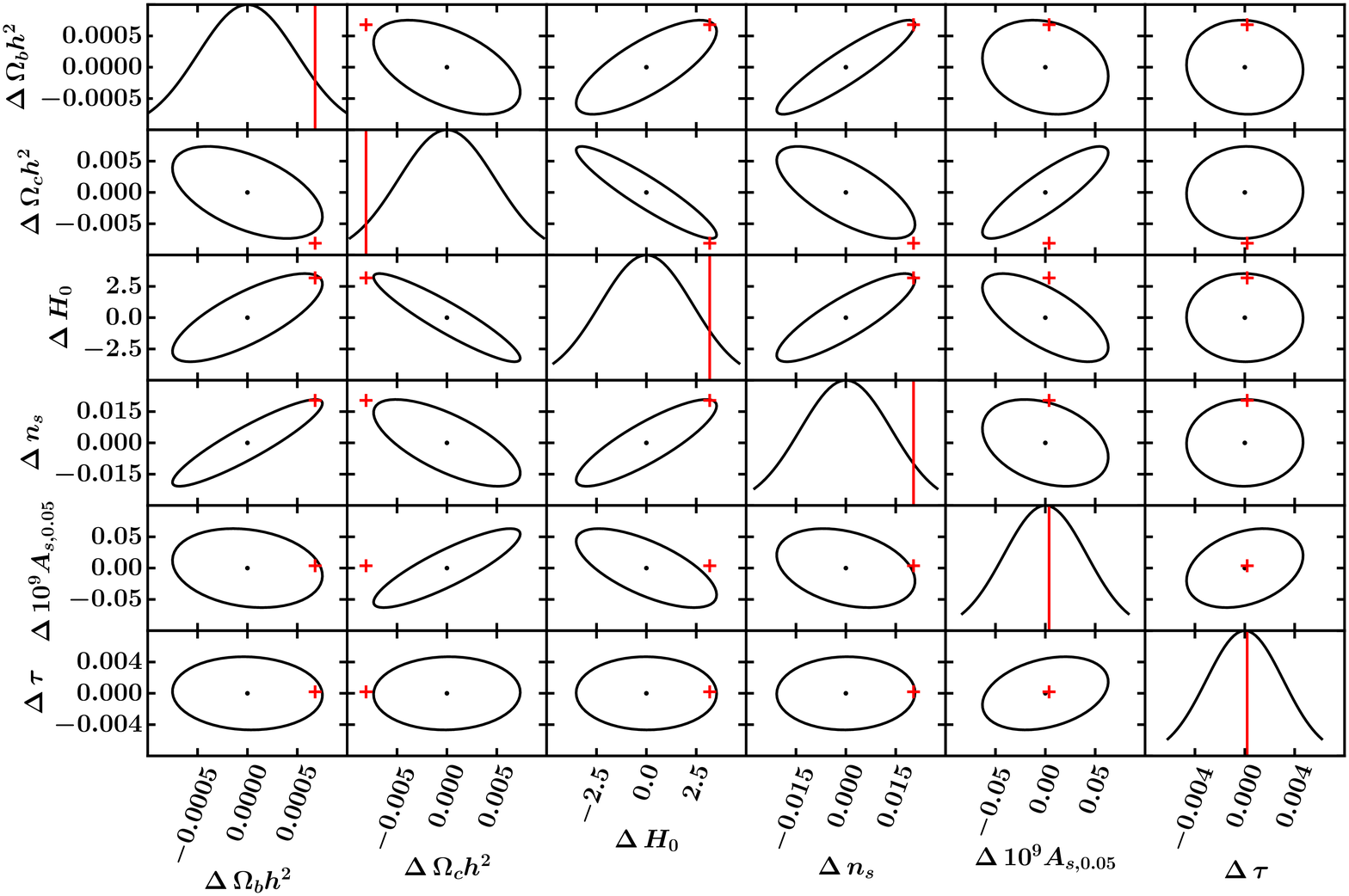}
\vspace{-0.15in}
\end{center}
\caption{A comparison of the 6 \lcdm\ model parameters derived from the \planck\ and \wmap\ data, respectively.  The vertical lines in the diagonal panels show the measured  difference for each parameter separately; the plus symbols in the off-diagonal panels show the measured parameter differences pairwise. The curves in the diagonal panels show the likelihood of the single parameter difference, taking into account the fact that both satellites observe a common sky, but employ different sky cuts, have different $l$ ranges, and have different instrument noise.  The ovals in the off-diagonal panels show the pair-wise covariance of the expected parameter differences.  Note that each single measured parameter is consistent within the individual uncertainties, but several pairs of parameter differences lie outside their 68\% CL ellipses, particularly the pair $A_{s,0.05}$ and $\Omega_ch^2$. Taken as a whole, the $\chi^2$ for the \Planck$-$\wmap\ parameter difference is 56 for 6 degrees of freedom, which has a probability to exceed of $\sim 3\times 10^{-10}$.  This indicates that the two experiments infer significantly different sets of \lcdm\ parameters.
\label{fig:planck_minus_wmap}}
\end{figure*}

We use the $6\times 6$ covariance matrix of parameter differences, with and without rescaling, to assess the consistency of the measured parameter difference.  The results are shown in Figure~\ref{fig:planck_minus_wmap}.  When we do not rescale the simulations at all we find $\chi^2 = 85.3$.  Upon rescaling each of the 12 standard deviations to match the published errors, we find $\chi^2 = 56$.  To assess possible systematic uncertainty in the scaling, we also sample the 12 parameter standard deviations uniformly in the interval between the published data value and the simulation result.  The resulting distribution of $\chi^2$ values is well approximated by a Gaussian with a mean of 69.2 and a standard deviation of 6.6.  The lowest $\chi^2$ observed in this test was $38.8$.  All of these values have extremely low probabilities to exceed (PTE): $\chi^2=38.8$ has PTE $< 10^{-6}$ for six degrees of freedom.  Our best estimate is $\chi^2=56$, which lies well within the extrema of 39 to 85.

We conclude that the differences between the two parameter measurements cannot be attributed to the effects of different sky cuts, multipole ranges, or instrument noise.  The most striking pair-wise discrepancy lies in the sub-space ($\Omega_c h^2$,$A_{s,0.05}$), where \planck\ observes a higher value of $\Omega_c h^2$ than \wmap, but a nearly identical value of $A_{s,0.05}$, causing the pair-wise difference to lie well outside the 68\% CL ellipse.  To probe this difference further, we marginalize $\chi^2$ over these two parameters, both individually and together, to assess the consistency of the remaining parameters; the results are summarized in Table~\ref{tab:p_chi2}.  For reference, the first row shows the high $\chi^2$ value for the full 6 parameter set.  Next, we show the results of marginalizing over $A_{s,0.05}$ and $\Omega_c h^2$: marginalizing over either single parameter drops $\chi^2$ to a reasonable value.   Marginalizing over any of the other 4 single parameters does not substantially reduce $\chi^2$.  Next we show that marginalizing over both parameters gives a low $\chi^2$ (but with a still-reasonable PTE).  Finally, we show that marginalizing over the 4 parameters \{$\Omega_b h^2$, $H_0$, $n_s$, $\tau$\} leaves a high $\chi^2 / \nu$ with a PTE of just 0.26\% for the pair \{$\Delta A_{s,0.05}$, $\Delta \Omega_c h^2$\}.  

The last column of Table~\ref{tab:param_comp} shows single parameter differences and their uncertainties, each after marginalizing over the other 5 cosmological parameters.  We note that they all individually agree, with the largest difference being $1.7\sigma$ for $\Omega_c h^2$; curiously, $A_{s,0.05}$ differs by only $0.1\sigma$.  Collectively, the \wmap\ and \planck\ best-fit parameters are discrepant by $\sim$$6\sigma$, with $\Omega_c h^2$ being the most discrepant single parameter.

Table~A.1 of \cite{planck/16:2013} compares cosmological parameters derived from \wmap\ data to parameters derived from the \planck\ data restricted to $l<1000$.  In that comparison, they find that $\Omega_c h^2$ is more consistent ($\Delta\Omega_c h^2 = 0.0005$)  and that the amplitude parameters differ by 1.2\%, but still not the 2.5\% one would naively expect from the spectrum ratios.  The remainder of the spectrum difference is accounted for by small changes in the remaining \lcdm\ parameters.  Thus, at least some of the parameter tension appears to arise from multipoles $l \gtrsim 1000$.

\capstartfalse
\begin{deluxetable}{lrrcr}
\tablecaption{Parameter consistency tests \label{tab:p_chi2}}
\tablewidth{0pt}
\tablehead{\colhead{Parameter(s) marginalized} & \colhead{$\chi^2$} & \colhead{$\nu$} & \colhead{$\chi^2/\nu$} & \colhead{PTE\tablenotemark{a}} }
\startdata
None                                 & 55.9 & 6 & 9.3  & $<10^{-6}$ \\
$A_{s,0.05}$                         & 6.3  & 5 & 1.3  & 0.28 \\
$\Omega_c h^2$                       & 5.2  & 5 & 1.0  & 0.39  \\
$A_{s,0.05}$, $\Omega_c h^2$         & 2.2  & 4 & 0.6 & 0.70 \\
$\Omega_b h^2$, $H_0$, $n_s$, $\tau$ & 11.9 & 2 & 6.0  & 0.0026
\enddata
\tablenotetext{a}{Probability to exceed, accounting for degrees of freedom,~$\nu$.}
\end{deluxetable}
\capstarttrue

\subsection{Discussion}\label{sec:param_discuss}

For full-sky experiments like \wmap\ and \planck, measurements of the angular power spectrum rely on calibrating the overall brightness scale to the CMB dipole.  This calibration is transferred to smaller angular scales by means of a window function, $W_l$, which mainly depends on the beam response, but it can also depend on the filtering applied to the time-ordered data (e.g. the \planck\ cosmic ray correction procedure).  Given the similarity of the \wmap\ and \planck\ dipoles, the differences between the two inferred power spectra are most likely explained by errors in the window functions of one or both experiments.

For \wmap, the window function is dominated by the beam response of the experiment.  In \S\ref{sec:wmap_beam} we review the steps that lead to the \wmap\ brightness calibration and window function and conclude that the power spectrum cannot be altered by more than a fraction of its current uncertainty of $\sim$0.6\% in amplitude ($\sim$1.2\% in power), and that the most likely sources of error would {\it increase} the power spectrum amplitude slightly.

Given its more sensitive instruments, the \planck\ window function must account for more effects than does \wmap's.  We do not have access to the data required to assess these effects, but in \S\ref{sec:fg_cln_compare} we argue that certain features in the sky map differences resemble recognized far sidelobe features in the \planck\ data.  We speculate that this may indicate that residual errors remain in the overall \planck\ window function. We note that \wmap\ was able to measure its beams in flight using 17 seasons of crisscrossed Jupiter observations, while the \planck\ observing pattern was more restricted, leading to a somewhat higher reliance on beam models.  Further, the HFI detectors saturate on Jupiter, so the HFI beam measurements must employ a combination of sky sources to cover the full dynamic range of the beam response.

To explore the sensitivity of cosmological parameters to \wmap\ and \planck\ beam errors, we perform the following exercise: we take the largest beam error mode from each experiment's likelihood function, and form a modified window function for each experiment.  For \planck\ we use the $143 \times 143$ GHz beam error provided with the \planck\ likelihood, since it covers the largest $l$ range; the \wmap\ likelihood provides a single set of beam error modes, so no choice is required.  For each case, we refit \lcdm\ parameters to the modified spectra.  The \wmap\ beam error couples primarily to $A_{s,0.05}$ and $\tau$, due to the steepness of the beam error rise at $l \sim 100$.  The \planck\
beam error couples to almost all of the parameters at a similar level, including $\Omega_c h^2$.  Based on this exercise, it seems unlikely that a \wmap\ dominant mode beam error would affect $\Omega_c h^2$, while a \planck\ beam error might.

\section{Conclusions}\label{sec:conclusions}

In this paper we report on consistency checks of the \planck\ and \wmap\ data.  At first glance, the \wmap\ and \planck\ data agree remarkably well, however, there exist small but significant differences between them.  We summarize our conclusions below.

\newcounter{mycounter}
\vspace*{0.15in}\stepcounter{mycounter}\noindent(\arabic{mycounter})
Examining the raw \planck\ $-$ \wmap\ difference maps $30-$Ka, $44-$Q, $70-$V and $100-$W (in thermodynamic temperature units after smoothing to a common resolution of $2^\circ$), we find that much of the CMB anisotropy signal cancels, but calibration differences between the two experiments leave residual dipole differences because the dipole is a relatively strong signal.  The [44,Q], [70,V] and [100,W] dipole differences are well aligned with the CMB dipole direction, with amplitudes of 0.3\%, 0.3\% and 0.5\% of the CMB dipole, respectively.  All are positive in the south and negative in the north, consistent with the \wmap\ maps being brighter than the \planck\ maps. These residuals are within the quoted uncertainties for both the \planck\ and \wmap\ data \citep{planck/03:2013, planck/08:2013, bennett/etal:2013} and are insufficient to explain the power spectrum differences at higher multipoles, by at least a factor of 2.  Since the effective frequencies for each pair of maps are not identical there is also imperfect cancellation of foregrounds in the [44,Q], [70,V] and [100,W] difference maps.

\vspace*{0.15in}\stepcounter{mycounter}\noindent(\arabic{mycounter})
After foreground template cleaning, the LFI$-$\wmap\ differences exhibit a quadrupole component higher than expected based on the different treatments of the kinematic quadrupole. The HFI 100 GHz $-$ \wmap\ W band difference map exhibits large-scale systematic structure that is not solely limited to a quadrupolar signature.  We identify this systematic structure as associated with the \planck\ map, possibly originating from beam sidelobe pick-up effects, either directly or indirectly. Note that the \planck\ Collaboration did not remove far sidelobe pickup from the delivered 100, 143, 217, or 353 GHz maps so some residual is expected.

\vspace*{0.15in}\stepcounter{mycounter}\noindent(\arabic{mycounter})
We form a template-independent, foreground-reduced difference map using an extension of the internal linear combination technique.  Again, we find features that are plausibly associated with residual \planck\ systematic errors. Lacking the full \planck\ time-ordered data we cannot fully address these features.

\vspace*{0.15in}\stepcounter{mycounter}\noindent(\arabic{mycounter})
Over a wide range of angular scales, the fluctuations measured by \wmap\ are $\sim$1.2\% brighter than measured by \planck\ ($\sim$2.5\% in power).  On the largest scale probed---the dipole moment---the experiments agree to within the uncertainties.  This agreement may persist up to multipoles of a few 10's, as seen in Figure 1, but residual foreground differences arising from bandpass differences make it difficult to draw definitive conclusions on these scales.  In model fits, the amplitudes of the first acoustic peak at $\ell=220$ differ by $\sim150\,\mu$K$^2$ ($\sim$2.7\%) between the two experiments.  Using simulations, we examine the best-fit value of $D_{220} \equiv l(l+1)C_{l}/(2\pi) |_{l=220}$ for an ensemble of \lcdm\ spectra fit to the \wmap\ and \planck\ data, and estimate the significance of this difference to be $3-5\sigma$, depending on the treatment of the window function uncertainties.
 
\vspace*{0.15in}\stepcounter{mycounter}\noindent(\arabic{mycounter})
We have re-examined the \wmap\ beam measurements and analysis seeking an explanation for the observed difference between the \planck\ and \wmap\ power spectra at intermediate angular scales.  We conclude that residual errors in characterizing the \wmap\ beam response cannot plausibly explain the $\sim$2.5\% power spectrum difference.

\vspace*{0.15in}\stepcounter{mycounter}\noindent(\arabic{mycounter})
We have evaluated the consistency of the cosmological parameters inferred from each experiment, taking into account the similarities and differences between them.  They observe the same sky but they cover different multipole ranges, employ different sky masks, and have different instrument noise.  We develop simulations to account for these effects and, for the first time to our knowledge, derive a $6 \times 6$ covariance matrix of parameter differences.  The 6 \lcdm\ parameters derived from the two experiments are individually consistent with each other, after marginalizing over the remaining 5 parameters.  The largest difference we find is $\Delta \Omega_c h^2 = -0.0081 \pm 0.0049$ (1.7$\sigma$).  However the full set of 6 \lcdm\ parameters are discrepant at the $\sim6\sigma$ level, with a $\chi^2=56$ for 6 degrees of freedom (PTE = $3 \times 10^{-10})$. We have estimated the uncertainty of this $\chi^2$ in \S\ref{sec:param_result}.

\vspace*{0.15in}\stepcounter{mycounter}\noindent(\arabic{mycounter})
The $\sim$6$\sigma$ \planck-\wmap\ parameter discrepancy is driven by two parameters: the the cold dark matter density, $\Omega_c h^2$, and the power spectrum amplitude, $A_{s,0.05}$.  The $\chi^2$ of the pair-wise difference \{$\Delta A_{s,0.05}$, $\Delta \Omega_c h^2$\} is 11.9 for two degrees of freedom, with a PTE of just 0.26\%.  Figure~\ref{fig:planck_minus_wmap} shows that $\Omega_c h^2$ and $A_{s,0.05}$ are somewhat degenerate over the range of multipoles probed by \wmap, and that the fits prefer to boost the value of $\Omega_c h^2$ rather than decrease $A_{s,0.05}$ to fit the lower \planck\ spectrum.   Marginalizing over either $A_{s,0.05}$ or  $\Omega_c h^2$ separately renders the remaining five parameters consistent.  Our analysis results are consistent with \citet{hazra/shafieloo:2014b}, but demonstrate that other parameter combinations provide comparable or statistically preferable fits.

\vspace*{0.15in}\stepcounter{mycounter}\noindent(\arabic{mycounter})
We explored the sensitivity of cosmological parameters to \wmap\ and \planck\ beam errors with a simple exercise: we constructed a modified window function for each experiment (\S\ref{sec:param_discuss}) and refit the \lcdm\ parameters to correspondingly modified power spectra.  We found that the dominant \wmap\ beam error coupled mostly to $A_{s,0.05}$ and $\tau$, while the dominant \planck\ beam error coupled to most of the parameters, including $\Omega_c h^2$.  

\vspace*{0.15in}
The disagreement between the \planck\ and \wmap\ data could signal a failure of the 6-parameter \lcdm\ model, and hence point to evidence for new physics.  However, it is of primary importance to first fully understand and account for systematic measurement errors such as those apparent in the sky map and power spectrum differences reported here and by the \Planck\ Collaboration.  We note that \citet{planck/31:2013} reports some changes in the \Planck\ window function relative to the 2013 versions used in this paper.  This update brings the \planck\ spectrum closer to \wmap's, but a $\sim$2\% discrepancy remains near the first acoustic peak.   Once this difference is resolved, it will be interesting to see if small differences in the \lcdm\ parameters persist.

\vspace*{0.15in}
This research was supported in part by NASA grant NNX14AF64G and by the Canadian Institute for Advanced Research (CIFAR).  We thank the \planck\ team for providing us with a copy of \citet{planck/31:2013} in advance of its publication.  We acknowledge use of the HEALPix \citep{gorski/etal:2005} and CAMB \citep{lewis/challinor/lasenby:2000} packages.  This research has made use of NASA's Astrophysics Data System Bibliographic Services.  We acknowledge the use of the Legacy Archive for Microwave Background Data Analysis (LAMBDA), part of the High Energy Astrophysics Science Archive Center (HEASARC). HEASARC/LAMBDA is a service of the Astrophysics Science Division at the NASA Goddard Space Flight Center.  We also acknowledge use of the \Planck\ Legacy Archive. \Planck\ is an ESA science mission with instruments and contributions directly funded by ESA Member States, NASA, and Canada.

\appendix
\section{Harmonic Conventions}\label{sec:conventions}

In this paper we quote coefficients for monopoles, dipoles, and quadrupoles
using real-valued spherical harmonics.
To make the numbers more readily understandable, we use basis functions that
peak at $\pm 1$ for the monopole and dipole.  Our monopole basis function is 1 
everywhere, and not $Y_{0,0} = 1/\sqrt{4\pi}$, so that when we quote a monopole
coefficient it is the monopole offset of the map, and not that offset divided by 
$\sqrt{4\pi}$.  Likewise, our $x$, $y$, and $z$ dipole basis functions are 
$\cos(\phi)\sin(\theta)$, $\sin(\phi)\sin(\theta)$, and $\cos(\theta)$, respectively, where
$\theta$ and $\phi$ are the usual co-latitude and longitude angular coordinates.
This means that a dipole-only map represented by 
$T(\theta,\phi) = a\, x(\theta,\phi) + b\, y(\theta,\phi) + c\, z(\theta,\phi)$
has a peak value of $\sqrt{a^2 + b^2 + c^2}$ on the sky.
This matches the conventions in the \wmap\ papers where 3.355 mK is the peak
value of the measured dipole on the sky \citep{hinshaw/etal:2009}.

For the quadrupole and higher modes, we switch to the usual orthonormalization 
convention for the spherical harmonics.  This is in an attempt to match the sign
conventions in Table~7 of \citet{hinshaw/etal:2007}\footnote{Errata for
Table 7: Numbers in table for $l=2,3$ and $m\ne 0$ do not have
$\sqrt{2}$ multiplied in.  
The $l<0$, $l=0$, and $l>0$ 
in the notes should be $m<0$, $m=0$, and $m>0$, respectively.
We should have $\tilde{a}_{lm} = \sqrt{2}\,\mathrm{Im}\,a_{l|m|}$
for $m<0$, using standard $a_{l m}$ values from Healpy.
Dipole normalization is for $x$,$y$,$z$ harmonics
that peak at $\pm 1$, not $\tilde{a}_{1,m}$ values, for numbers in table.
Error bars on $l,b$
for $(d,l,b)$ triple should be switched in footnote $a$.}
and Table~III of \citet{deoliveira-costa/etal:2004a}.

A software package such as Healpy is able to calculate spherical harmonic
coefficients $a_{l m}$ so that the temperature field can be described as
\be
T(\theta, \phi) = \sum_{l=0}^{l_{\rm max}}\sum_{m = -l}^l a_{l m} Y_{l m}(\theta, \phi)
\ee
See for example Section 5.2 of 
\citet{varshalovich/moskalev/khersonskii:1988} for a definition
of the spherical harmonics $Y_{l,m}(\theta,\phi)$, and Section 5.13
of that book for their formulas written out explicitly.
For real-valued fields T, we have the relation
\be
a_{l,-m} = (-1)^m a_{l,m}^*.
\ee
which means the $a_{l m}$ coefficients for negative $m$ are completely determined by those for positive $m$.
Healpy only stores $a_{l m}$ for $m \ge 0$, to save space.
For ease of use, we define real-valued coefficients as follows.
\be
\tilde{a}_{l m} = \left\{
\begin{array}{r@{\quad\mathrm{if}\quad}l}
\sqrt{2}\; \mathrm{Im}[a_{l|m|}] & m < 0 \\
a_{l m} & m = 0 \\
\sqrt{2}\; \mathrm{Re}[a_{l m}] & m > 0 
\end{array}
\right.
\ee
For an alternative definition of real-valued spherical
harmonics, see
\citet{chisholm:1976} and \citet{blanco/florez/bermejo:1997},
although Chisholm adds an extra ``c'' or ``s'' index to the harmonics
instead of allowing negative values of $m$, so he does not specify
a natural ordering of the coefficients.

\begin{figure}[ht]
\begin{center}
\includegraphics[width=6.5in]{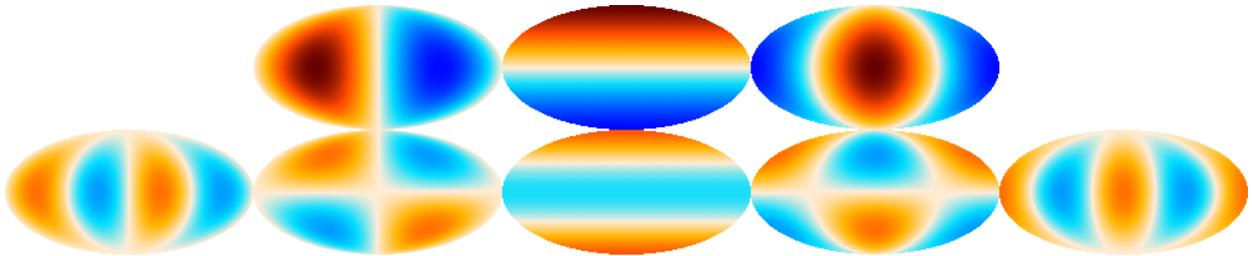}
\end{center}
\caption{
The real-valued dipole and quadrupole modes used in this paper.
In the top row are the $y$, $z$, and $x$ dipole modes, with the order
chosen to align with the quadrupole modes in the bottom row.
The dipole modes reach their maximum value of $+1$ (colored orange-brown)
in the directions of the positive $y$, $z$, and $x$ axes, respectively.
From this it is clear that our Mollweide projection uses the astronomical convention
that the observer is inside the sphere: $\theta$ increases downwards, and $\phi$
increases when moving right to left in this image.
On the bottom row,
real-valued quadrupole modes $\tilde{Y}_{2,m}$, defined in 
Appendix~\ref{sec:conventions}, are ordered from $m=-2$ on the left to $m=2$ on the right.
Color scale is $\pm 1$ for each image.  Note that the dipole basis functions have a larger
amplitude than the quadrupole basis functions due to our different choices of normalization.
Also note that the $m=-1$ quadrupole mode
is prominent in various \Planck/\wmap\ differences, in galactic coordinates.
\label{fig:dipole_quadrupole}
}
\end{figure}

To determine the basis functions corresponding to these 
$\tilde{a}_{l m}$ coefficients, consider a real-valued map
that contains only two modes for a specific $l$, and for $m$ and
$-m$, with $m>0$.
\be
T = a_{l m}Y_{l m} + a_{l,-m} Y_{l,-m}
\ee
Since the temperature field is real-valued, we have
\be
T = a_{l m}Y_{l m} + a_{l m}^* Y_{l m}^* 
= 2\, \mathrm{Re}[a_{l m}Y_{l m}]
= 2\, \mathrm{Re}[a_{l m}]\mathrm{Re}[Y_{l m}]
- 2\, \mathrm{Im}[a_{l m}]\mathrm{Im}[Y_{l m}]
\ee
This reduces to
\be
T = \tilde{a}_{l m}\cdot \sqrt{2}\,\mathrm{Re}[Y_{l m}]
+ \tilde{a}_{l, -m}\cdot (-\sqrt{2}\,\mathrm{Im}[Y_{l m}])
= \tilde{a}_{l m}\tilde{Y}_{l m}
+ \tilde{a}_{l,-m}\tilde{Y}_{l,-m}
\ee
where we have defined real-valued basis functions
$\tilde{Y}_{l m}$ as:
\be
\tilde{Y}_{l m} = \left\{
\begin{array}{r@{\quad\mathrm{if}\quad}l}
-\sqrt{2}\,\mathrm{Im}[Y_{l|m|}]
= (-1)^m \sqrt{2}\,\mathrm{Im}[Y_{l,-|m|}] & m < 0 \\
Y_{l m} & m = 0 \\
\sqrt{2}\; \mathrm{Re}[Y_{l m}] & m > 0 
\end{array}
\right.
\ee
This allows us to expand a general real-valued temperature field
in real-valued harmonics up to some maximum multipole $l_{\rm max}$
as
\be
T(\theta, \phi) = \sum_{l = 0}^{l_{\rm max}} \sum_{m=-l}^l 
\tilde{a}_{l m} \tilde{Y}_{l m}(\theta, \phi).
\ee
Our dipole $x,y,z$ and quadrupole $\tilde{Y}_{2,m}$ modes are shown in Figure~\ref{fig:dipole_quadrupole}.

To relate the normalization convention for the $\tilde{Y}_{1 m}$ modes
to our preferred dipole modes, we have:
\be
x(\theta,\phi) = -\sqrt{\frac{4\pi}{3}} \tilde{Y}_{1,1}(\theta,\phi)
\ee
\be
y(\theta,\phi) = \sqrt{\frac{4\pi}{3}} \tilde{Y}_{1,-1}(\theta,\phi)
\ee
\be
z(\theta,\phi) = \sqrt{\frac{4\pi}{3}} \tilde{Y}_{1,0}(\theta,\phi)
\ee

Our new basis of $\tilde{Y}_{l m}$ real-valued harmonics still 
has a similar normalization to the usual $Y_{l m}$ harmonics,
\be
\int\!\!\!\!\int \tilde{Y}_{l m} \tilde{Y}_{l'm'}\;d\cos\theta\; d\phi 
= \delta_{l l'}\delta_{mm'},
\ee
so we can quote the power in the quadrupole as the usual $C_l$ value, 
\be
C_2 = \frac{1}{5} \sum_{m=-2}^2 (\tilde{a}_{2,m})^2 \quad \rm{in}\;[\mu\rm{K}^2].
\ee

We also briefly discuss the kinematic quadrupole, which is of the form $Y_{2,0}$ and is already real-valued, in an appropriately
rotated coordinate system.  Since  $Y_{2,0}(\theta,\phi) = \frac14\sqrt{\frac5\pi}(3\cos(\theta)^2-1)$,
the peak-to-trough amplitude for this harmonic is $\frac34\sqrt{\frac5\pi} = 0.946175$.
Ignoring the frequency-dependent effects discussed in \cite{planck/27:2013},
the peak-to-trough amplitude of the kinematic quadrupole is $T_{\rm CMB}(v/c)^2 \approx 4.1\, \mu$K, where $v$ is the velocity
of the observer with respect to the CMB rest frame, and $c$ is the speed of light.  
If the peak-to-trough distance is 4.1 $\mu$K, the $a_{2,0}$ value is $4.1/0.946175 \approx 4.3\,\mu$K.
Since the kinematic quadrupole has no $m\ne 0$ power in our rotated reference frame, we have only one term in 
our sum for the kinematic quadrupole power: 
$C_2 = \frac15 a_{2,0}^2 \approx 3.8\,\mu{\rm K}^2$.

\section{Extended ILC method}\label{sec:eilc_app}

To extend our ILC method to look at systematic differences between experiments,
we follow much the same algebra as in the appendix of \citet{eriksen/etal:2004d}.
As they do, we minimize the variance of a linear combination of maps, but
we use the different constraints that the sum of weights for the first set of maps must 
be $+1$, and the sum of the weights for the second set of maps must be $-1$.

The variance that we minimize is still their Equation A1:
\be
\mathrm{Var}(T) = \mathbf{w}^T \mathbf{C} \mathbf{w}
\ee
where $\mathbf{w}$ is a vector of weights and $\mathbf{C}$ is
the covariance matrix of the maps $T_i$ (which must all be in the same units) given by their Equation A2.
\be
C_{ij} = \langle \Delta T_i \Delta T_j \rangle
= \frac{1}{N_{\rm pix}} \sum_{p=1}^{N_{\rm pix}}
(T^i(p) - \bar{T}^i)(T^j(p) - \bar{T}^j)
\ee
We wish to minimize their Equation A3
\be
f(\mathbf{w}) = \sum_{ij}w_i C_{ij} w_j
\ee
subject to two constraints:
\be
g(\mathbf{w}) = \sum_{i=1}^{n_1} w_i = 1,
\ee
and
\be
h(\mathbf{w}) = \sum_{i=n_1+1}^{n_2} w_i = -1.
\ee
The vector $\mathbf{w}$ has indices 1 through $n_1$ corresponding
to the bands of the first experiment, and indices $n_1+1$ through $n_2$ 
corresponding to the bands of the second experiment.  Any additional templates
which should be removed from the ILC (and which presumably contain no CMB) can be added in with 
indices above $n_2$; these have unconstrained weights.  If the vector $\mathbf{w}$
has $n$ indices, then indices $n_2+1$ to $n$ correspond to extra templates.
Eriksen's Equation A5 is now modified for our constraints:
\be
\nabla f(\mathbf{w}_0) = \lambda_1 \nabla g(\mathbf{w}_0) + \lambda_2 \nabla h(\mathbf{w}_0)
\ee
We convert this, in addition to the two direct constraints on the weights above, 
into the $n+2$ by $n+2$ matrix equation written out in blocks:
\be
\label{eq:lagrange}
\left(
\begin{array}{cc}
2 \mathbf{C} & -\Updownarrow \\
\Leftrightarrow & \mathbf{0}
\end{array}
\right)
\left(\begin{array}{c} \mathbf{w} \\ \lambda_1 \\ -\lambda_2 \end{array} \right)
=
\left(\begin{array}{c} \mathbf{0} \\ 1 \\ -1 \end{array} \right),
\ee
where $\Leftrightarrow$ is the $2\times n$ matrix of 1s and 0s that encodes the constraints
on the weights, and $\Updownarrow\, =\, \Leftrightarrow^T$. 
For example, if we have 5 \wmap\ bands so that $n_1=5$, and we have 7 more \Planck\ bands
so that $n_2 = 12$, and we have three templates so that $n=15$, then we use
\be
\Leftrightarrow \, =
\left(
\begin{array}{ccccccccccccccc}
1 & 1 & 1 & 1 & 1 & 0 & 0 & 0 & 0 & 0 & 0 & 0 & 0 & 0 & 0 \\
0 & 0 & 0 & 0 & 0 & 1 & 1 & 1 & 1 & 1 & 1 & 1 & 0 & 0 & 0 \\
\end{array}
\right).
\ee
The matrix in Equation~\ref{eq:lagrange} 
can be constructed directly and inverted to solve for $\mathbf{w}$,
$\lambda_1$, and $\lambda_2$.  Since the matrix is so small ($17\times 17$ in the above
example), the matrix inversion is a rapid operation.

If the matrix in Equation~\ref{eq:lagrange} is singular, 
then there is probably a linear dependence among the maps or templates $\Delta T_i$,
making the matrix $\mathbf{C}$ singular as well.
One way to remedy this is to remove one or more of the maps or templates from the calculation until 
the matrix $\mathbf{C}$ becomes invertible again, allowing a unique solution to Equation~\ref{eq:lagrange}.
This is only likely to be a problem if a map has been duplicated in the calculation of $\mathbf{C}$;
the independent noise from maps at different frequencies tends to make the maps linearly independent.

\section{Simplifying the \textsl{Planck} likelihood}\label{sec:simplify_planck}

\Planck\ has released a set of cross power spectra, for $100 \times 100$, $143 \times 143$, $143 \times 217$, and $217\times 217$ GHz, extractable from their released likelihood code.
These use $l$ ranges of 50--1200, 50--2000, 500--2500, and 500--2500, respectively.  \Planck\ provides the measured values of these
power spectra and their $7104\times 7104$ element covariance matrix.

Four spectra are useful when studying foregrounds in our galaxy,
but for cosmology we only want one: the spectrum of the primordial CMB, characterized by the frequency dependence of
a blackbody of constant temperature.
Therefore, we find the best fit power spectrum given the above information.  This is the spectrum $\hat{C}$ which 
is the best fit to the below matrix equation, explicitly written out in blocks:
\be
\left(\begin{array}{cccc}
\mathbb{I}_{450} & 0_{450\times 701} & 0_{450\times 800} & 0_{450\times 500} \\
0_{701\times 450} & \mathbb{I}_{701} & 0_{701\times 800} & 0_{701\times 500} \\
\mathbb{I}_{450} & 0_{450\times 701} & 0_{450\times 800} & 0_{450\times 500} \\
0_{701\times 450} & \mathbb{I}_{701} & 0_{701\times 800} & 0_{701\times 500} \\
0_{800\times 450} & 0_{800\times 701} & \mathbb{I}_{800} & 0_{800\times 500} \\
0_{701\times 450} & \mathbb{I}_{701} & 0_{701\times 800} & 0_{701\times 500} \\
0_{800\times 450} & 0_{800\times 701} & \mathbb{I}_{800} & 0_{800\times 500} \\
0_{500\times 450} & 0_{500\times 701} & 0_{500\times 800} & \mathbb{I}_{500} \\
0_{701\times 450} & \mathbb{I}_{701} & 0_{701\times 800} & 0_{701\times 500} \\
0_{800\times 450} & 0_{800\times 701} & \mathbb{I}_{800} & 0_{800\times 500} \\
0_{500\times 450} & 0_{500\times 701} & 0_{500\times 800} & \mathbb{I}_{500} \\
\end{array}\right)
\left(\begin{array}{c}
\hat{C}_{50-499} \\
\hat{C}_{500-1200} \\
\hat{C}_{1201-2000} \\
\hat{C}_{2001-2500} \\
\end{array}\right)
=
\left(\begin{array}{c}
C_{50-499}^{100\times 100} \\
C_{500-1200}^{100\times 100}  \\
C_{50-499}^{143\times 143} \\
C_{500-1200}^{143\times 143}  \\
C_{1201-2000}^{143\times 143}  \\
C_{500-1200}^{143\times 217}  \\
C_{1201-2000}^{143\times 217}  \\
C_{2001-2500}^{143\times 217} \\
C_{500-1200}^{217\times 217}  \\
C_{1201-2000}^{217\times 217}  \\
C_{2001-2500}^{217\times 217} \\
\end{array}\right),
\ee
where $0_{n\times m}$ is a matrix of zeros with $n$ rows and $m$ columns, 
and $\mathbb{I}_{n}$ is an $n\times n$ identity matrix.  Also, $\hat{C}_{n-m}$
is a column vector with $m-n+1$ elements, corresponding to a best fit power spectrum with
multipole $l$ values ranging from $n$ to $m$, inclusive.
The frequency dependent cross spectra $C^{x\times y}_{n-m}$ are cross spectra of $x$ GHz and $y$ GHz maps
also with multipole $l$ values ranging from $n$ to $m$, inclusive.
More compactly, this is
\be
A \hat{C} = C,
\ee
where $A$ is a $7104 \times 2451$ matrix of 1s and 0s, $\hat{C}$ is the 2451 element column vector for which we 
are solving, and $C$ is a 7104 element column vector of measured $C_l$ values at various frequencies.
If $N$ is the covariance matrix of $C$, then the linear least squares solution is:
\be
\label{eq:appendix_lstsq}
\hat{C} = (A^T N^{-1} A)^{-1} A^T N^{-1} C.
\ee

Since the measured power spectrum at different frequencies has different levels of foregrounds, we choose to consider
foreground-related spectra to be completely unconstrained modes.   This means making sure that $N^{-1}$ acting 
on a pure-foreground-spectrum mode zeros it.  Let $F$ be a matrix with foreground spectra in columns.
This assumes that the \Planck\ foregrounds can be constructed by a linear combination of templates.  Since there 
are some nonlinear features to the \Planck\ likelihood foreground model 
(for example, the beam mode template and calibration factors multiply the other foreground templates), 
we linearize around reasonable values of these calibration and foreground parameters.
To include these foreground templates into Equation~\ref{eq:appendix_lstsq},
we replace $N$ with $N + \infty F F^T$, which means replacing $N^{-1}$ with $(N + \infty F F^T)^{-1}$.
This should be done with the Sherman-Morrison-Woodbury formula, since we already have $N^{-1}$.
\be
N' = N + \sigma_f^2 F F^{T}
\ee
\be
(N')^{-1} = N^{-1} - N^{-1} F (\sigma_f^{-2} \mathbb{I} + F^T N^{-1} F)^{-1} F^T N^{-1}
\ee
We take the limit $\sigma_f \rightarrow \infty$ and find
\be
(N')^{-1} = N^{-1} - N^{-1} F (F^T N^{-1} F)^{-1} F^T N^{-1}.
\ee

Note that this makes the matrix $A^T (N')^{-1} A$ singular, so we have to take the pseudo-inverse when finding the linear
least squares solution.  This is equivalent to projecting the foreground modes out of our final spectrum.

\lastpagefootnotes

We explicitly construct $\hat{N} = (A^T (N')^{-1} A)^{-1}$, since this is the (singular) 
covariance matrix of our power spectrum $\hat{C}$.
With the pseudo-inverse, foreground modes are zeroed by this covariance 
matrix\footnote{A CMB power spectrum fluctuation that precisely matches the shape of a foreground 
template is therefore not measurable.},
but it is otherwise accurate and useful.

For reference, we expand the solution for the foreground-removed single power spectrum $\hat{C}$  
in terms of the original matrices $A$, $N$, and $F$:
\be
\hat{C} = (A^T [N^{-1} - N^{-1} F (F^T N^{-1} F)^{-1} F^T N^{-1}]  A)^{-1} A^T [N^{-1} - N^{-1} F (F^T N^{-1} F)^{-1} F^T N^{-1}] C.
\ee

\section{Polynomial estimation of $C_l^{\rm th}$}\label{app:pico}

We randomly sample 10,000 sets of six parameters, using a uniform distribution over the following intervals:
$\Omega_b h^2 \in [0.020, 0.025]$, 
$\Omega_c h^2 \in [0.10, 0.15]$, 
$H_0 \in [60, 80]$ km s$^{-1}$ Mpc$^{-1}$, 
$n_s \in [0.8, 1.1]$,
$A_s \in [2.0 \times 10^{-9}, 2.5 \times 10^{-9}]$ for $k_0=0.05$ Mpc$^{-1}$, and
$\tau \in [0.03, 0.20]$.
These intervals are chosen to encompass current parameter measurements, while being small enough to allow a fourth-order polynomial to fit the $C_l^{\rm th}$ values over the given range.  For each set of parameters, we calculate the lensed spectrum using CAMB, as described below.  For each multipole in the range $2 \le l \le 2500$, we fit $l(l+1)C_l^{\rm th}/2\pi$ to a fourth-order polynomial, using the first 9000 members of the sampled parameters; we save the remaining 1000 members for testing, as described below.  With 6 variables, a fourth-order polynomial has 210 terms\footnote{With six variables, the 0th order part has 1 term; the 1st order part has 6 terms; the 2nd order part has 21 terms;
3rd has 56 terms; and 4th has 126 terms.},
so it is well constrained when fit to 9000 samples.  We check the accuracy of the polynomial prediction for CAMB
with the remaining 1000 samples.  We find the maximum absolute value of the error (for any $l$) to be 
$|\Delta l(l+1)C_l^{TT}/2\pi| < 4.5 \mu{\rm K}^2$, 
with the median absolute error being much smaller, at 0.09 $\mu{\rm K}^2$.
Evaluating this 4th order polynomial is much faster than running CAMB itself, since the polynomial
can be evaluated in only about a million floating point operations (210 times 2500 times 2, plus a few),
which incidentally is fast enough to be almost real-time in a web browser window if implemented in JavaScript.
Another useful benefit is that the quartic polynomial is differentiable to machine precision, unlike
CAMB, which is only differentiable to the accuracy needed for MCMC parameter estimation, usually about 0.1\%.
This increased differentiability is very useful for several different optimization methods.
We note that this fit was inspired by and follows the same philosophy as that of the PICO 
code\footnote{\url{https://sites.google.com/a/ucdavis.edu/pico/home}} \citep{fendt/wandelt:2007,fendt/wandelt:2007b},
and PICO has been shown to work for a much larger range of parameters by using multiple polynomial fits.

When computing with CAMB, we use standard values for parameters not in our six-parameter model.
For example, we set the curvature, $\Omega_k$, and tensor-to-scalar ratio, $r$, both to be zero, and we use 
{\tt massless\_neutrinos~=~3.046} and {\tt massive\_neutrinos~=~0}.
We use the July 2013 version of CAMB, with the following also set in the parameter file:
{\tt l\_max\_scalar~=~3000}, {\tt k\_eta\_max\_scalar~=~7500},
{\tt temp\_cmb~=~2.7255},   {\tt helium\_fraction~=~0.24},
{\tt high\_accuracy\_default~=~T},  {\tt{accuracy\_boost~=~1}}, 
{\tt l\_accuracy\_boost~=~1}, and {\tt l\_sample\_boost~=~1}.

We run CAMB individually for each set of parameters, for simplicity, although further speed-ups could
be obtained in the initial 10,000 runs of CAMB 
by computing $C_l$ for several different values of the parameters affecting the primordial $P(k)$, 
namely $A_s$ and $n_s$, within a single run of CAMB.  This can be done by only modifying the CAMB
input parameter {\tt .ini} file.

\bibliographystyle{larson_apj}
\bibliography{wmap,planck}

\end{document}